\documentclass{article} 

\usepackage{amsmath}
\usepackage{amstext}
\usepackage{amsfonts}
\usepackage{amssymb,amscd}
\usepackage{amsopn}
\usepackage{graphicx,multirow,makecell}
\usepackage{amssymb,amscd,color,cite}

\usepackage{fullpage}

\definecolor{redm}{rgb}{0.858824, 0.00784314, 0.00784314}
\definecolor{bluem}{rgb}{0.24, 0.6, 0.33692}
\definecolor{brownm}{rgb}{0.6, 0.426641, 0.24}
\definecolor{greenm}{rgb}{0.24, 0.6, 0.33692}
\definecolor{bgreenm}{rgb}{0.145098, 0.435294, 0.384314}

\begin{document}
\begin{center}
{\Large Minimal R{\'e}nyi--Ingarden--Urbanik entropy
  of multipartite quantum states}

\vspace{0.4cm}

Marco Enr\'iquez $^{1,2}$, Zbigniew Pucha\l{}a $^{3,4}$ and Karol {\.Z}yczkowski $^{4,5}$

\vspace{0.3cm}
{\footnotesize
$^1$ Departamento de F{\'i}sica, Cinvestav, AP 14-740, 07000 M{\'e}xico~DF, M{\'e}xico. \\
$^2$ SEPI-UPIITA, Instituto Polit\'ecnico Nacional, Av. IPN No. 2580, Col. La Laguna Ticom\'an, \\ C.P. 07340 M\'exico DF, M\'exico \\
$^3$ Institute of Theoretical and Applied Informatics, Polish Academy of Sciences, \\ Ba\l{}tycka 5, 44-100 Gliwice, Poland\\
$^4$ Smoluchowski Institute of Physics, Jagiellonian University, ul. Reymonta 4, \\ PL-30-059 Krak{\'o}w, Poland\\
$^5$ Center for Theoretical Physics, Polish Academy of Sciences, Aleja Lotnik{\'o}w 32/46, \\ PL-02-668 Warsaw, Poland}

\vspace{0.5cm}

June 15, 2015

\end{center}
 
\noindent {\bf Abstract}. We study the entanglement of a pure state of a composite 
   quantum system consisting of several subsystems with $d$ levels each.
  It can be described by the R\'enyi--Ingarden--Urbanik entropy $S_q$ of a decomposition
of the state in a product basis, minimized over all local unitary transformations.
In the case $q=0$ this quantity becomes a function of the rank of the tensor
representing the state, while in the limit $q \to \infty$ the entropy becomes related
to the overlap with the closest separable state and the geometric measure of 
entanglement. For any bipartite system the entropy $S_1$ coincides 
with the standard entanglement entropy. We analyze the distribution of the minimal
entropy for random states of three and four--qubit systems. 
In the former case the distributions of $3$--tangle is studied and 
some of its moments are evaluated, while in the latter case
we analyze the distribution of the hyperdeterminant.
The behavior of the maximum overlap of a three--qudit system with the closest separable 
state is also investigated in the asymptotic limit.

\bigskip


\section{Introduction}

After more than twenty years of intensive research
entanglement of the pure states of bipartite quantum systems
is rather well understood \cite{HHH09,BZ06}
for two subsystems of an arbitrary dimension $d$.
 In this case any pure state can be represented in a product basis by a matrix
of coefficients of order $d$, 
and its standard singular values decomposition allows one to
reveal entanglement.

On the other hand, in the case of systems composed of $n \ge 3$ subsystems
 the problem becomes much more complicated, as the state is represented by a
tensor of size $d$ and $n$ dimensions. Even though several important results 
were obtained, especially in the case of three \cite{Aci00,Su01,WEJSSVW11}
and four qubits \cite{LT03,Le06,Gou10,VES11,SS12},
 and several  measures of entanglement in such systems were proposed
\cite{OH10,HS10,GS10,RHZ11}, 
it is fair to say that the complete understanding of the phenomenon
of entanglement in multipartite systems is still missing.
 
To characterize entanglement of a quantum state of a bi-partite system
it is natural to analyze the degree of mixing of the reduced density matrix.
For instance, making use of von Neumann entropy, one arrives at one of the most often used
measures: the entropy of entanglement \cite{Ben96}.
It is often convenient to apply for this purpose also the generalized entropy of R{\'e}nyi
\cite{BZ06} or some other kinds of entropy.
 
The aim of the present work is to propose a possible generalization
of this quantity for the case of multipartite systems, for which a pure state
is represented by a tensor. Furthermore, we would like to make a connection
with the geometric measure of entanglement, which depends on the distance of the state considered 
to the closest separable state \cite{Bro01,Zyc02,Wei03}.

Following the papers of Parker and Rijmen  \cite{Par01}  
and Bravi \cite{Bra03}
we suggest to analyze the entropy of decomposition of a
quantum state in a product basis, sometimes called {\sl Ingarden -- Urbanik entropy} \cite{Ing62}, minimized over all local unitaries.
This quantity
can be generalized in the sense of R{\'e}nyi. Interestingly, to establish a direct link 
with the geometric measure of entanglement \cite{Bro01,Wei03} it is sufficient
to consider the R{\'e}nyi -- Ingarden -- Urbanik (RIU) entropy
and send the Renyi parameter $q$ to infinity. 

Even though the approach advocated here is applicable for arbitrary
composite quantum systems, for concreteness we concentrate the
majority of this work for the case of three and four qubits.
It is demonstrated that there is no a single pure state for which the minimal RIU entropy 
is the largest for all values of the Renyi parameter $q$.
Investigating the problem for selected values of $q$ 
we identify certain pure states, which are conjectured to maximize this
particular measure of entanglement.

Furthermore, we make also use of a statistical approach
to analyze the distribution of the hyperdeterminant and minimal RIU entropy
for random quantum pure states.
They are distributed with respect to the unique unitarily invariant Haar measure
on the space of pure quantum states, the complex projective space,
${\mathbb{C}}P^{N-1}$, where the total
dimension of the complex Hilbert space is $N=d^n$. 
Analyzing systems composed of three subsystems of an arbitrary
dimension $d$ we obtain a bounds for the geometric measure of entanglement
for generic states of such a system.

This paper is organized as follows.
In Section 2 we introduce the RIU entropy for a pure state of a multipartite system, 
while in Section 3 different techniques of tensor decomposition are reviewed.
In Section 4 we present results obtained for three--qubit system, while analogous results for
4 qubits are presented in Section 5. A more general case of 
three subsystems consisting of an arbitrary number of levels is
discussed in Section 6. Computations of the moments of the distribution of 3--tangle
and a derivation of the bound for the geometric measure of
 entnaglement are relegated to the Appendix.


\section{Minimal R{\'e}nyi--Ingarden--Urbanik entropy}
\label{smriu}

Consider a quantum state describing a system consisting of $n$
subsystems, with $d$ levels each,
$|\psi\rangle \in {\cal H}_N={\cal H}_d^{\otimes n}$.
Working in an arbitrary product basis one
can represent such a state by a $n$--index tensor, 

\begin{equation} 
\vert \psi\rangle = \sum_{i_1=1}^d \dots \sum_{i_n=1}^d
  C_{i_1,i_2,\dots, i_n}  \vert i_1\rangle \otimes \vert i_2\rangle \otimes \dots \otimes \vert i_n\rangle 
  \quad \mbox{with }C_{i_1,i_2,\dots, i_n} \in \mathbb{C}.
\label{tensor2}
\end{equation}
The standard normalization condition, $\langle \psi|\psi\rangle =1$,
implies that 
\begin{equation} 
\sum_{i_1=1}^d \dots \sum_{i_n=1}^d \vert C_{i_1,i_2,\dots, i_n}\vert ^2=1.
\end{equation}
It will be convenient to introduce a multi-index $\mu = (i_1,i_2,\dots, i_n)$,
where $\mu$ can be identified with the set $\{1, \dots, N=n^d\}$,
and use a shorter notation $p_{\mu}=| C_{\mu}|^2=| C_{i_1,i_2,\dots, i_n}|^2$.
Hence $p(|\psi\rangle)=(p_1,\dots p_N)$ represents an $N$--point probability vector $\vec p$,
which can be characterized by the R{\'e}nyi entropy
$S_q(\vec p)=\frac{1}{1-q} \log \Big( \sum_{\mu=1}^N p_{\mu}^q \Bigr)$.
For $q\to 1$ this quantity reduces to the standard Shannon entropy
$S(p)= - \sum_{\mu=1}^N  p_{\mu} \log p_{\mu}$, which in the context 
of the decomposition of the state $|\psi\rangle$,
is called the {\sl Ingarden--Urbanik} entropy \cite{Ing62, Car00, Spe00}
and written $S^{\rm IU}(|\psi\rangle)=S\bigl(p(|\psi\rangle)\bigr)$.
In will be convenient to use natural logarithms throughout this paper, 
written $\log 2\approx 0.693$.

The product basis 
$ \vert i_1\rangle \otimes \vert i_2\rangle \otimes \dots \otimes \vert i_n\rangle$
is determined up to a local unitary transformation,
$U_{\rm loc}=V_1 \otimes V_2 \otimes \cdots \otimes V_n$,
where an unitary matrix $V_j$ acts on the $j$--th subsystem.
As the Ingarden--Urbanik entropy of the decomposition 
depends explicitly on the choice of the product basis
it is natural to analyze the optimal value, minimized over 
the set of local unitaries \cite{Car00,Spe00}.

We shall study a more general case of the R\'enyi entropies $S_q$
with R\'enyi parameter $q\ge 0$. For any multipartite
state $\vert \psi\rangle \in {\cal H}_d^{\otimes n}$
we define the  minimal {R{\'e}nyi--Ingarden--Urbanik} (RIU) entropy,
\begin{equation} 
S^{\rm RIU}_q \left(\psi \right) \ := \
 \min_{U_{\rm loc}}   S_q \left[ p( U_{\rm loc} \vert \psi\rangle )\right],
\label{RIU}
\end{equation}
where the minimum is taken over entire set of local unitary transformations.

{\bf Proposition 1}. For any $N$ qudit state,
$\vert \psi\rangle \in {\cal H}_d^{\otimes n}$, its  
 minimal RIU entropy is bounded from above,
$S^{\rm RIU}_q (\psi) \le \log R_{\rm max}$,
where 
\begin{equation}
R_{\rm max}= d^n - nd(d-1)/2.
\label{rmax}
\end{equation}

This statement follows directly from the work of
Carteret, Higuchi and Sudbery \cite{Car00}, 
who showed that performing a local unitary transformation 
$U_{\rm loc}=U_1 \otimes \dots \otimes U_n$,
one can always find a product basis
 such that the decomposition~(\ref{tensor2}) contains no more 
than $R_{\rm max}$ terms. A suitable choice of $n$ unitary matrices
of size $d$ allows one to bring $d(d-1)/2$ elements of the tensor $C$ 
to zero. Therefore out of all its $d^n$ entries 
at least $nd(d-1)/2$ can be always set to zero. This fact 
can be also formulated as the following statement

{\bf Proposition 2}. For any $n$ qudit state,
$\vert \psi\rangle \in {\cal H}_d^{\otimes n}$,
represented by a tensor $C$ as in Eq. (\ref{tensor2}),
its {\sl tensor rank} $R$ is bounded by $R_{max}=d^n - nd(d-1)/2$,
where $R$ is the minimal number $r$ of terms
in its decomposition involving arbitrary coefficients $f_{\nu}$,
\begin{equation} 
 C_{i_1,i_2,\dots, i_n} = \sum_{\nu=1}^r f_{\nu} a_{i_1}^{\nu} b_{i_2}^{\nu}
  \dots z_{i_n}^{\nu} .
\label{t_rank}
\end{equation}
\medskip

Observe that in the case of the R\'enyi entropy of order zero, $q=0$, 
one has $S^{\rm RIU}_0(\psi)=\log R$
where $R$ is the tensor rank of the tensor $C(\psi)$.
The quantity $\log R$ is known as the {\sl Schmidt measure} characterizing entanglement
of the multipartite pure state \cite{Eis01,Hei04}, while the rank $R$ was used to determine
probabilistic conversion of three qubit pure state \cite{Chi08}.
Since the R\'enyi entropy is in general a non-increasing function of the parameter $q$,
one gets the general bound 
\begin{equation} 
S^{\rm RIU}_q ( \vert \psi\rangle) \le  \log R(\vert \psi\rangle ) \le \log R_{\rm max}.
\label{RIU0}
\end{equation}
For instance, in the case of a three--qubit system, ($n=3$ and $d=2$),
the bound (\ref{rmax}) gives
$R_{\rm max}=8-3=5$ in agreement to the five--term standard form of
a three--qubit pure state by Acin {\em et al} \cite{Aci00}.

Besides the case a) $q=0$, corresponding to the tensor rank of $\vert \psi\rangle$,
we shall also consider some other particular cases of the definition (\ref{RIU}).

b) $q=1$. The minimal IU entropy $S^{\rm IU}_1 (\vert \psi\rangle)$ determines the minimal
 information gained by environment after performing a projective von--Neumann
 measurement of the pure state $\vert \psi\rangle \langle \psi \vert$ in an 
arbitrary product basis \cite{Bra03}.

c) $q=2$. The minimal decomposition entropy $S^{\rm RIU}_2 (\vert \psi\rangle)$ characterizes the maximal 
  purity $\sum_{\mu} p^2_{\mu}=e^{-S_2}$ of the probability vector $p_{\mu}$ 
associated to the outcomes of a projective  measurement in a product basis
  and is accessible in a coincidence experiment with two copies of the
 multipartite state $\vert \psi\rangle$. This quantity was used by Parker and Rijmen \cite{Par01}
 to analyze multipartite entanglement in the context of coding theory. 
\medskip

d) $q=\infty$. In the limiting case $q\to \infty$ the minimal RIU entropy 
gives  $S^{\rm RIU}_{\infty} (\vert \psi\rangle)= - \log \lambda_{\rm max}$,
where the largest component of the vector $p_{\mu}$ reads
$\lambda_{\rm max}=\max \vert \langle \psi \vert \chi_{\rm sep}\rangle \vert ^2$.
The maximum is taken over the set of all separable states, 
$\vert \chi_{\rm sep} \rangle = U_{\rm loc} \vert 0 \cdots 0\rangle$,
so $\lambda_{\rm max}$ is a decreasing function of the
Fubini--Study distance to the closest separable state \cite{Bro01, Zyc02}, 
$D_{\rm FS}= \arccos( \sqrt{\lambda_{\rm max}})$, 
and its function called the  geometric measure of entanglement \cite{Wei03},
$E_G(\psi)= 1  - \lambda_{\rm max}=1-\exp[- S^{\rm RIU}_{\infty}(\vert \psi\rangle)]$.
\medskip

In the case of bipartite systems ($n=2$) one can use the standard singular
value decomposition of the matrix $C$ to show that the probability vector
$p_{\mu}$ coincides with the vector of Schmidt coefficients $\lambda_j$
determining the Schmidt decomposition \cite{BZ06},
$\vert \psi \rangle = \sum_{j=1}^d \sqrt{\lambda_j} \vert j\rangle_A \otimes \vert j\rangle_B$,
where $\sum_{j=1}^d \lambda_j =1$. This observation leads to 

{\bf Proposition 3}. For any bipartite state,
$\vert \psi\rangle \in {\cal H}_d \otimes  {\cal H}_d$, 
its  minimal RIU entropy coincides with the 
R\'enyi entropy of entanglement, so in the case $q=1$ 
one arrives at the standard entropy of entanglement
\begin{equation} 
S_1^{\rm RIU}(\psi)=E(\vert \psi\rangle) =  - \sum_{j=1}^d  \lambda_j \log \lambda_j 
 = S(\lambda) .
\label{entan_entr}
\end{equation}

\section{Tensor decompositions}\label{stendec}

Singular value decomposition (SVD) of the matrix containing the
expansion coefficients of a biparite pure state
plays a crucial role in evaluating the minimal RIU entropy. 
However, in the case of multipartite systems,
and quantum states described by higher order tensors,
this decomposition is not directly applicable.

In this section we consider two generalizations of SVD, 
the higher order singular value decomposition (HO-SVD)~\cite{Del00}
 and the Parallel Factor model (PARAFAC)~\cite{Bro97}. 
Both decompositions were developed in the framework of  
Principal Component Analysis   \cite{JMZ15}  and have found several
applications in signal processing, numerical linear algebra, graph analysis,
and numerical analysis \cite{Kol09}. Throughout this section we will consider
such schemes to evaluate some bounds for the minimal RIU entropy and the
geometric measure of entanglement. 

Let us define a tensor $C$ as a multidimensional or $n$-way array of numbers,
so it can be identified with an element of $\mathbb{C}^{d_1 d_2 \ldots d_n}$.
Accordingly, a matrix and a vector are a 2-way and 1-way tensors, respectively.
Note that such a space is linear.
Given two tensors $A,B \in \mathbb{C}^{d_1 d_2 \ldots d_n}$ their inner product
is inherited form the linear space $\mathbb{C}^{d_1 d_2 \ldots d_n}$ and
defined as follows
\begin{equation}
 \langle A,B \rangle = \displaystyle \sum_{i_1,i_2,\ldots,i_n} {\overline A_{i_1,i_2,\ldots,i_n} B_{i_1,i_2,\ldots,i_n}},
\end{equation}
where the over-bar denotes the complex conjugation. 
The corresponding induced norm is the Frobenius norm \cite{Kol09, Del00},
 written  $\Vert{A}\Vert := \sqrt{\langle A,A\rangle}$. 
It implies the Frobenius distance, $d(A,B) := \Vert A-B \Vert$. Note that 
that the coefficients of the state (\ref{tensor2}) can be arranged in a tensor $C\in\mathbb{C}^{d^n}$ 
and conversly, a given tensor in such a space defines a certain 
pure state $|\psi\rangle$  in the Hilbert space ${\cal H}_d^{\otimes n}$.
%

\subsection{Higher Order Singular Value Decomposition}

Let $C\in \mathbb{C}^{d_1 d_2 \ldots d_n}$ be an $n$-way tensor. We define the $k$-th unfolding $C_{(k)}$ as the matrix of size $d_k \times (d_{k+1} d_{k+2} \cdots d_n d_1 d_2\cdots d_{k-1})$ that contains the tensor element $C_{i_1,i_2,\ldots,i_n}$ in the entry $(i_k,j)$, with
\begin{equation*}
 j=1+\sum_{\substack{\ell=1\\ \ell\neq k}}^n (i_\ell-1)J_\ell, \qquad \qquad \mbox{with } J_\ell =\prod_{\substack{m=1\\ m\neq k}}^{l-1} d_m.
\end{equation*}
The $k$-mode product of a matrix $U^{(k)}\in \mathbb{C}^{d_k \times d_k}$ with $C$ is defined (element-wise) as
\begin{equation}\label{kprod}
 (U^{(k)} C)_{i_1, i_2,\ldots , i_{k-1}, i_k^\prime,i_{k+1},\ldots,i_n}=
 \sum_{i_k} C_{i_1, i_2,\ldots , i_{k-1}, i_k,i_{k+1},\ldots,i_n} u^{(k)}_{i_k^\prime,i_k}.
\end{equation}
The higher order singular value decomposition allows  one
to construct a tensor $A$, called {\it co-tensor} \cite{Kol09},
of the same dimension than $C$ such that
\begin{equation}\label{ctensor}
 A = U^{(1)}\otimes U^{(2)}\cdots \otimes U^{(n)} C,
\end{equation}
where each $U^{(k)}$ acts according to (\ref{kprod}) and any two sub-tensors $A_{i_k=p}$ and $A_{i_\ell=q}$, with $p$ and $q$ fixed, are orthonormal
\begin{equation}
 \langle A_{i_k=p}, A_{i_k=q} \rangle = \delta_{p,q} \left[\sigma_p^{(k)}\right]^2.
\end{equation}
The numbers $\sigma_p^{(k)}$ are called the $k$-mode singular values of $C$, they are non-negative and fulfill $\sigma_p^{(k)}\ge \sigma_q^{(k)}$ for all $p<q$. Such a decomposition is accomplished by taking each $U^{(k)}$ in (\ref{ctensor}) as the matrix of left singular vectors of $C_{(n)}$.
Note that finding the SVD of the $k$-th unfolding of the coefficients tensor of (\ref{tensor2}) is equivalent to diagonalize the reduced density matrix of the $k$-th party of the system. Moreover, according to equation (\ref{ctensor}) the co-tensor $A$ of $C$ defines 
a state $\vert \psi \rangle_{\rm HO-SVD}$ that is LU equivalent to (\ref{tensor2}). 
In this manner,  Liu {\it et al} have recently proposed an entanglement classification based on the study of HO-SVD and the local symmetries of the multipartite states \cite{Liu12}.

\subsection{Parallel Factor decomposition}

The idea of expressing a tensor as a sum of rank-one tensors was applied in several contexts. 
For instance, in psychometrics it is known as {\sl canonical decomposition} (CANDECOMP),
while in the brain imaging analysis it is referred to as
{\sl parallel factor decomposition} (PARAFAC) -- see \cite{Bro97} and references therein. 
 For other applications of this decomposition see also \cite{Kol09}. 

The PARAFAC is a decomposition of $n$-way 
tensor $C\in \mathbb{C}^{d_1 d_2 \ldots d_n}$ into a sum of rank-one tensors, 
\begin{equation}\label{parafac1}
 C
 = \sum_{k=1}^R \lambda_k ~ U^{(1)}_{i_2=k}\circ U^{(2)}_{i_2=k}\circ \cdots U^{(n)}_{i_2=k},
\end{equation}
where $U^{(\ell)}_{i_2=k}$ denotes the $k$-th column of a matrix of size $d_\ell\times R$,
with $\ell=1,\ldots,n$ and $\lambda$ is a vector of size $R$.
The symbol $\circ$ represents the dyadic product of vectors \cite{Enr13a}. In terms of the components, the PARAFAC decomposition reads
\begin{equation}
 C_{i_1,i_2,\ldots,i_n}
 = \sum_{k=1}^R \lambda_k ~ u_{i_1,k}^{(1)}u_{i_2,k}^{(2)}\cdots u_{i_n,k}^{(n)}.
\end{equation}
In practice, we are interested in an approximation given by sum of $r$ rank-one tensors
\begin{equation}
 C \simeq C^{\rm PARAFAC} 
 = \sum_{k=1}^r \lambda_k ~ U^{(1)}_{i_2=k}\circ U^{(2)}_{i_2=k}\circ \cdots U^{(n)}_{i_2=k},
\end{equation}
Usually a least squares approximation (\ref{parafac1}) is concern, in which case one
has to minimize the quantity $d_P=\Vert C - C^{\rm PARAFAC}\Vert$.
Since the problem is not linear, usually we obtain an approximate solution only,
accomplished through the alternating least squares (ALS) algorithm.
In this work we use the algorithm provided by Nion and De Lathauwer~\cite{Nio08}
to compute the PARAFAC decomposition of a 3-way complex tensor. In the case of $r=1$ the state
\begin{equation} 
\vert \psi\rangle_{\rm P} = \sum_{i_1=1}^d \dots \sum_{i_n=1}^d
  C_{i_1,i_2,\dots, i_n}^{\rm PARAFAC}  |i_1\rangle \otimes |i_2\rangle \otimes \dots |i_n\rangle,
\label{tensor3}
\end{equation}
is fully separable. Hence, the quantities $d_P$ and
$\lambda_p = \vert \langle \psi \vert \psi_{\rm P} \rangle\vert^2$
are bounds for the geometric measure of entanglement $E_G$ and the
separable state maximum overlap $\lambda_{\rm max}$, respectively.
\section{Three qubits}\label{sthrqub}

In this section we analyze pure states of a system
consisting of three qubits: $n=3$ and $d=2$.
The exact values of the minimal RIU entropy are found for the states $W$ and
$GHZ$ for particular values of the R\'enyi parameter $q$. Generic properties
of the minimal RIU entropy are also discussed. Furthermore, some moments of the
3--tangle $\tau$ are computed and the corresponding distribution is analyzed.

\subsection{Minimal decomposition entropy}\label{ssmind}

In the case of a  three--qubit system
any pure state  $\vert \psi\rangle \in {\cal H}_2^{\otimes 3}={\cal H}_8$
can be represented in the five--terms decomposition of Acin et al. \cite{Aci00},
\begin{equation} 
\vert \psi\rangle = a_1 \vert 000\rangle + a_2 \vert 001\rangle +a_3 \vert 010\rangle +a_4 \vert 100\rangle +a_5 \vert 111\rangle ,
\label{acin1}
\end{equation}
where $\sum_{i=1}^5 \vert a_i \vert ^2=1$ and four coefficients can be chosen
to be real. Observe that selecting $a_1=a_5=1/\sqrt{2}$ and neglecting others,
one obtains the state
$\vert GHZ\rangle=(\vert 000\rangle + \vert 111\rangle)/\sqrt{2}$,
while setting  $a_2=a_3=a_4=1/\sqrt{3}$ one has  
$\vert W\rangle=(\vert 001\rangle + \vert 010\rangle + \vert 100\rangle)/\sqrt{3}$.
As the number of terms in these states cannot be reduced
by any local unitary transformation
their tensor ranks are equal to two and three, respectively, 
so that $S^{\rm RIU}_0(GHZ)=\log  2$ and $S^{\rm RIU}_0(W)=\log 3$. 
It is possible to write a state $\vert A_5 \rangle$
with all coefficients equal, $a_i=1/\sqrt{5}$
for which $S^{\rm RIU}_0(A_5)=\log 5$. In the same form, we define
$\vert A_4 \rangle = (\vert 000 \rangle+\vert 010 \rangle + \vert 001\rangle+\vert 111 \rangle)/2$
such that $S_0^{\rm RIU} (A_4) = \log 4$.

\begin{figure}[htbp]
 \centering
\includegraphics[scale=0.75]{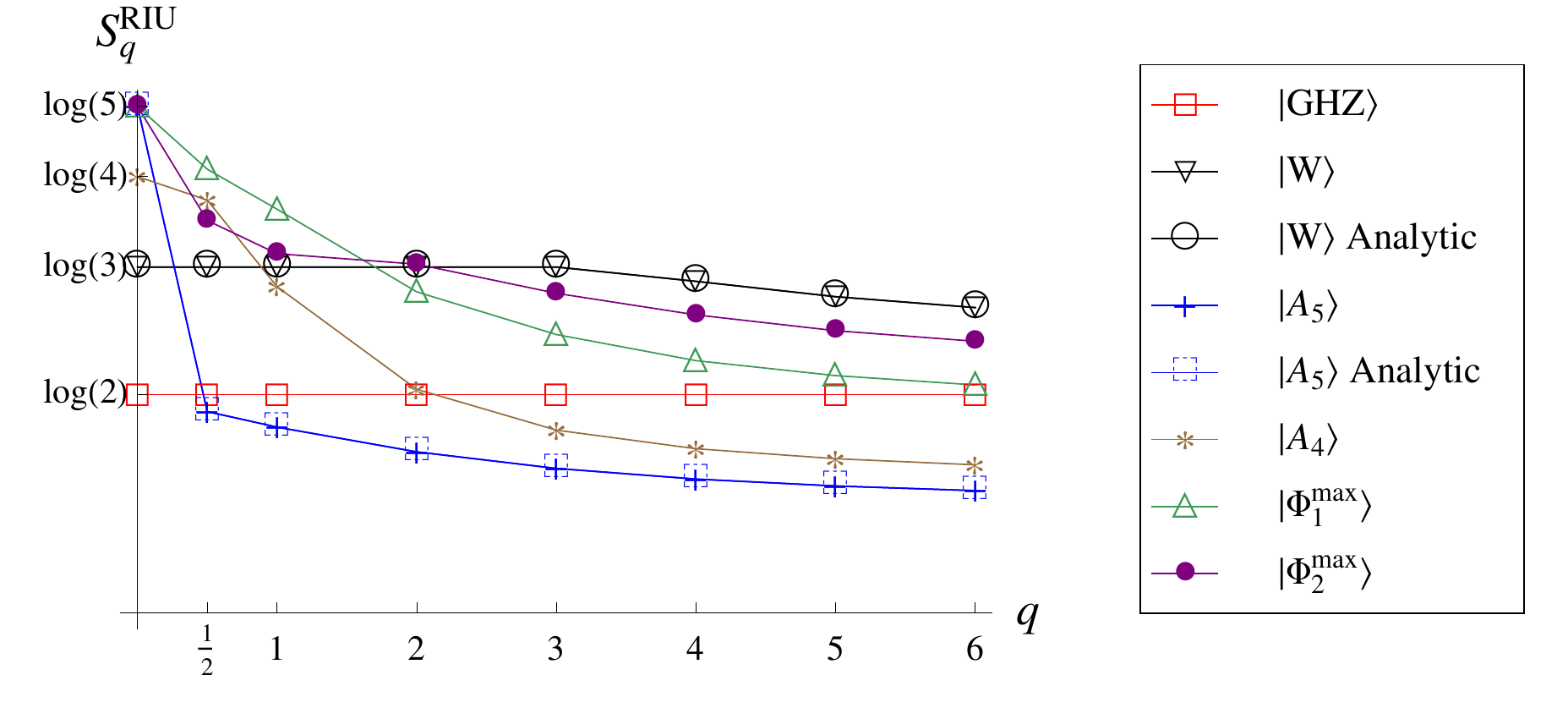}
\caption{(Color online)  Minimal RIU entropy obtained numerically for several
three qubit states as function of the R\'enyi parameter $q$.  
Comparison with analytical results available for states invariant with respect to 
permutation confirms accuracy of the numerical procedure.
}
\label{sqst}
\end{figure}

For any general value of the R\'enyi parameter $q$, we are not aware
of any constructive procedure which gives the exact value of minimal RIU entropy (\ref{RIU}).
However, for permutation invariant states we can follow the general scheme of 
calculating the maximum overlap with the closest separable state. For such
states it was first conjectured and later proven that in order to obtain the
maximal overlap it is enough to take the product state to be a tensor product
of the same single-party real state \cite{Hub09,Hay09,Wei10}. In this spirit,
to minimize the minimal RIU entropy for permutation invariant states the product
of local unitary matrices $U_{\rm loc}$ in (\ref{RIU}) will be taken as
\begin{equation}\label{ulocpi}
 U_{\rm loc} = U(p)^{\otimes 3}, \quad {\rm with }\quad U(p) = \left(\begin{array}{cc} \sqrt p & \sqrt{1-p}\\[1ex] -\sqrt{1-p} & \sqrt p \end{array} \right), \quad p\in [0,1].
\end{equation}
This task can be done easily in some special cases, i.e.
$S_q^{\rm RIU}( {\rm GHZ}) =\log 2$ for any $q$ and $S_1^{\rm RIU}({\rm W}) =\log 3$.  
To evaluate the expression (\ref{RIU}) for an arbitrary state, we perform a
random walk over the space of unitary matrices. Fig.~\ref{sqst} shows 
 the minimal RIU entropy obtained numerically
for some particular states and different values of $q$.
 In the case of the state $|W\rangle$ and $|A_5\rangle$ one can compare the value obtained by
this procedure with the analytical results.

\begin{figure}[tbp]
\centering
 \begin{tabular}{ccc}
 \includegraphics[scale=0.52]{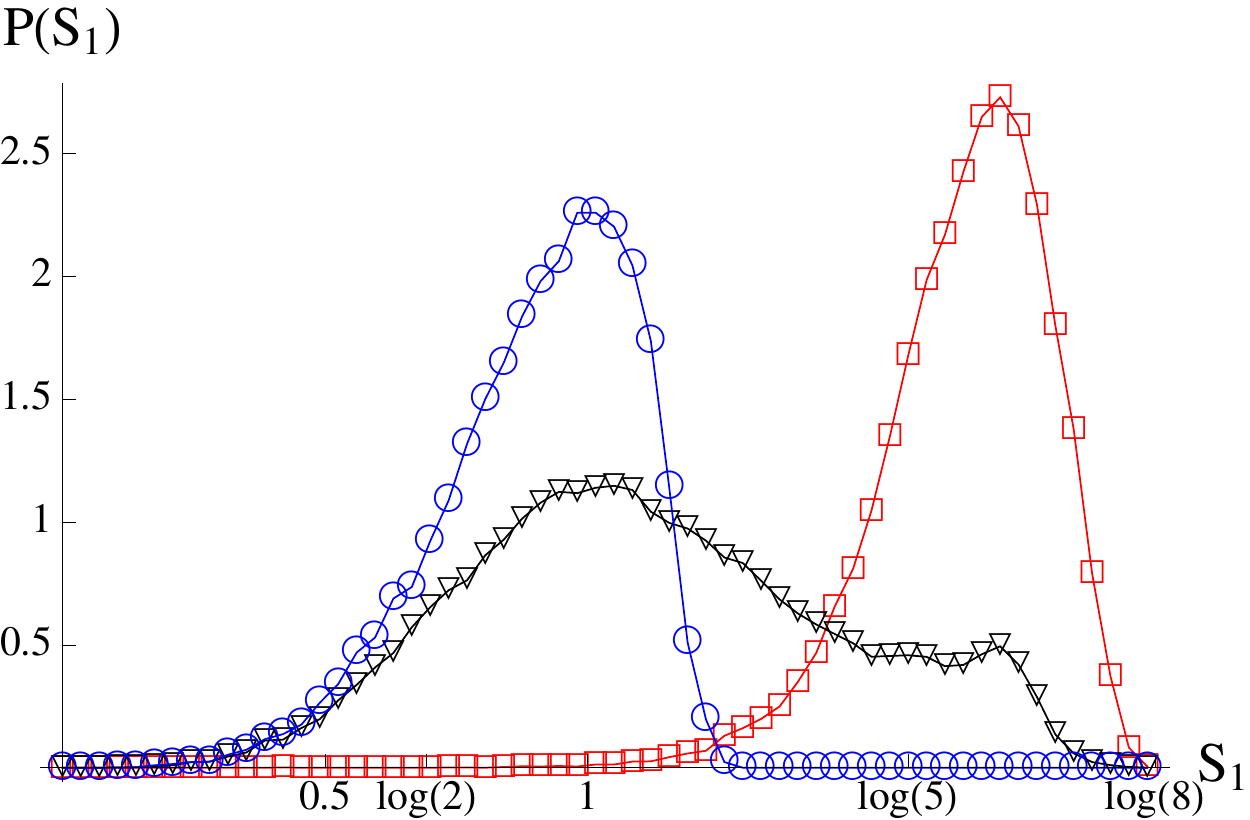}& \hspace*{0.5cm}& \includegraphics[scale=0.5]{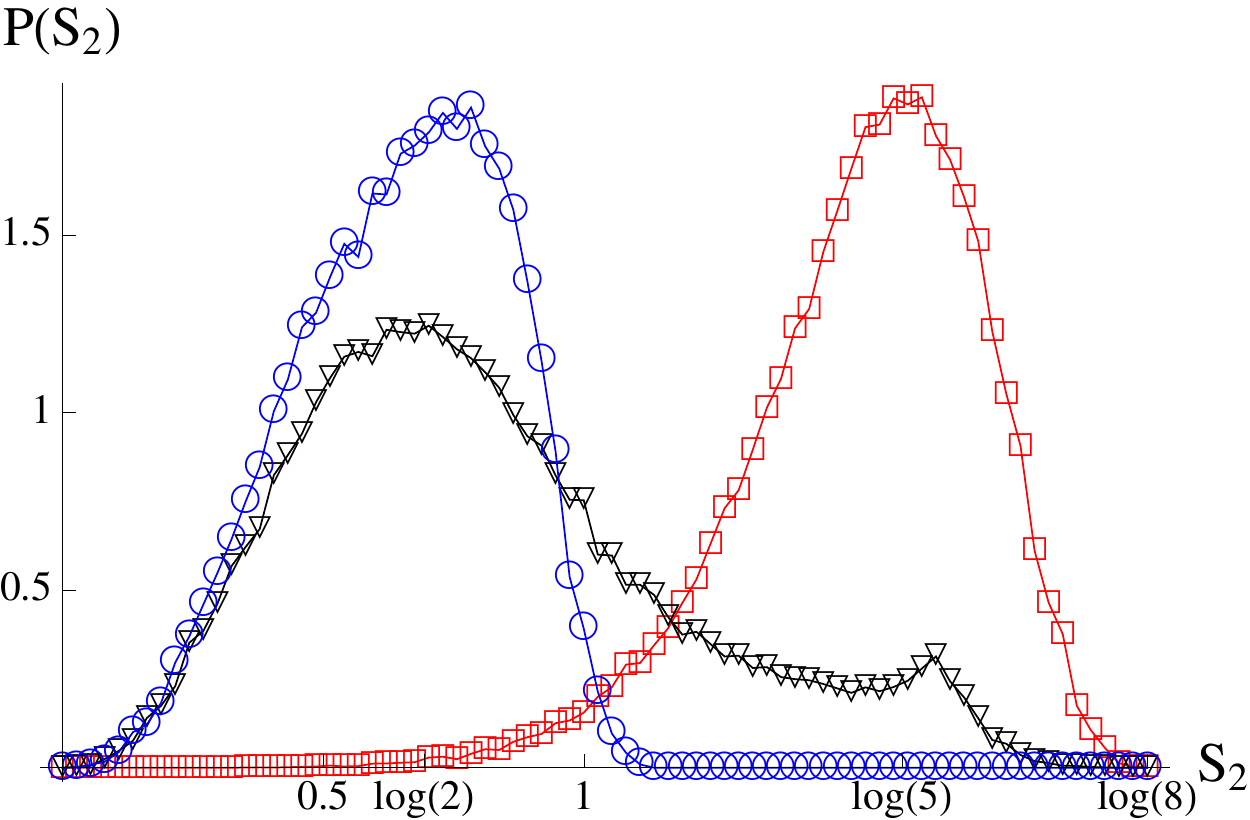} \\[1ex]
 a) & & b) \\[1ex]
\includegraphics[scale=0.55]{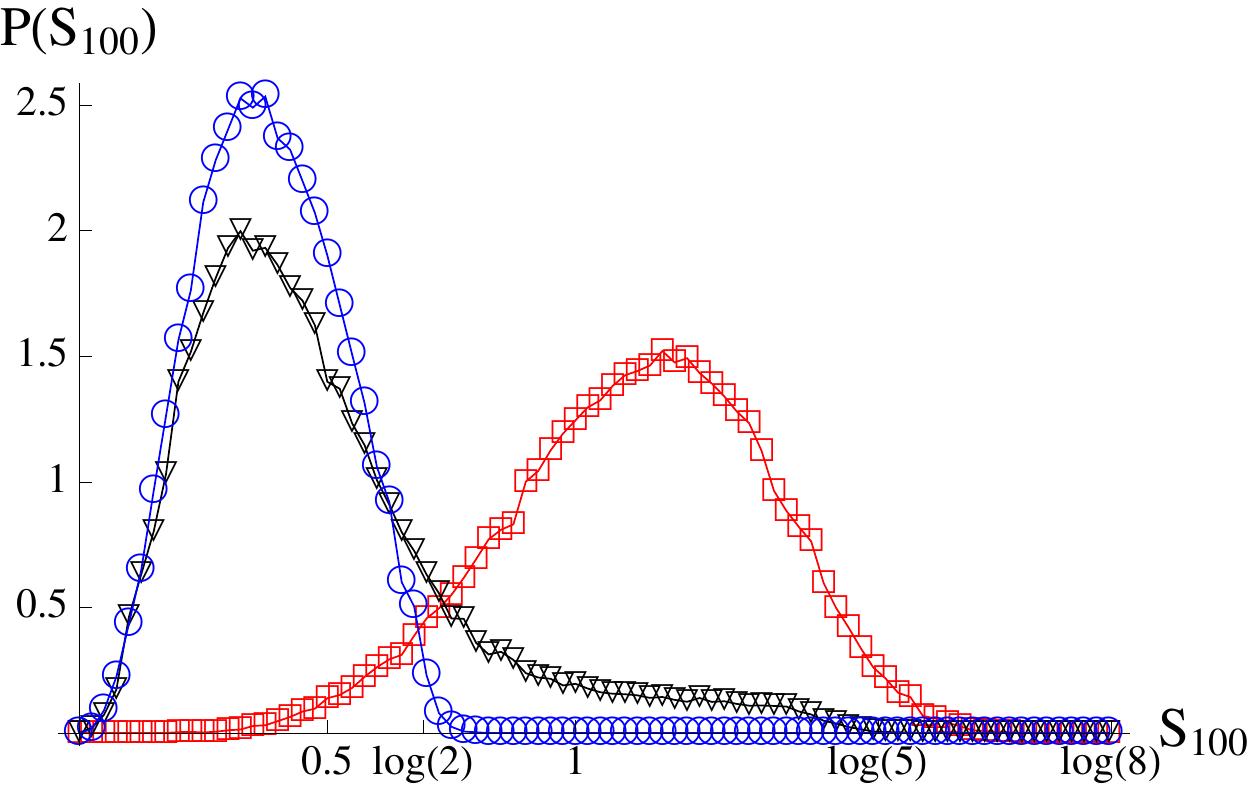} & & \includegraphics[scale=0.51]{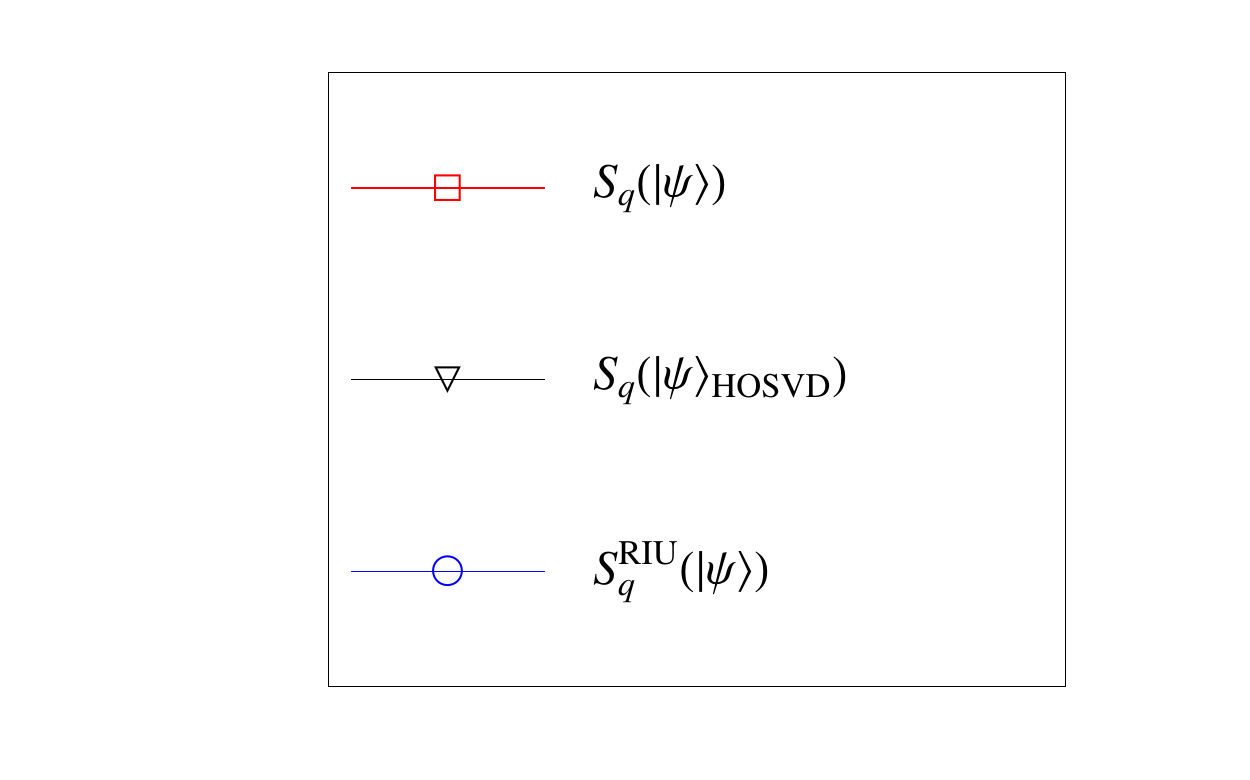}\\[1ex]
 c) & {} & {}
 \end{tabular}
\caption{(Color online) Distributions of the R\'enyi entropy of random probability vector
describing a generic three-qubit state ({$\color{red} \square$}), the R\'enyi entropy of the corresponding co-tensor ($\triangledown$) and the estimation of minimal RIU entropy ({$\color{blue} \circ$}) for (a) $q=1$, (b) $q=2$ and (c) $q=100$.}
\label{sqst3qr}
\end{figure}

\begin{figure}[tbp]
\centering
 \includegraphics[scale=0.64]{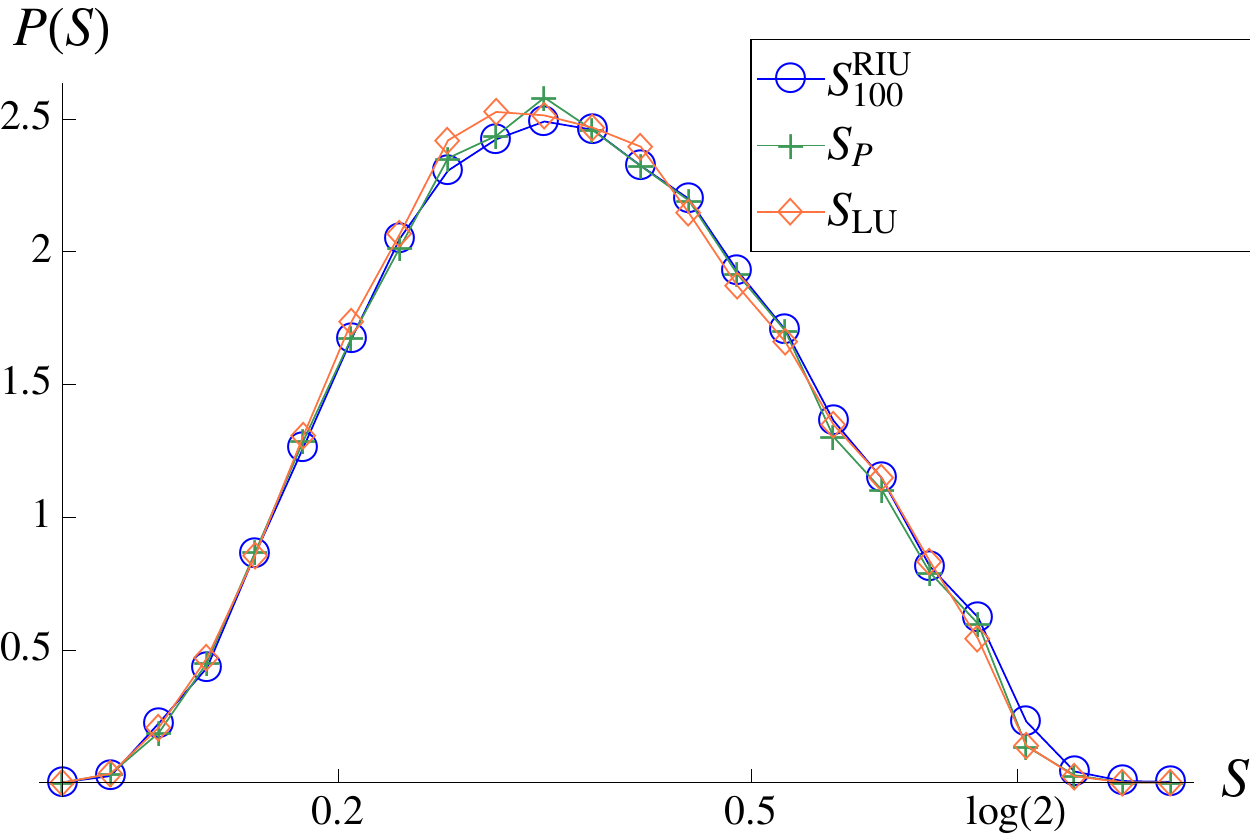}
 \caption{(Color online) Comparison between the distributions of $S_{100}^{\rm RIU}$ ({$\color{blue} \circ$}), $S_{\rm P}=- \log \lambda_{\rm max}^{\rm PARAFAC}$ (${\color{greenm} +}$) and $S_{\rm LU}=-\log \lambda_{\rm max}^{\rm LU}$ (${\color{black} \Diamond}$) for three--qubit random states.} 
\label{csqst3qr}
\end{figure}

In the following we are concerned with typical properties of the minimum value of RIU entropy. 
First, we compute the value of the R\'enyi entropy of a
random state, then the same quantity is evaluated in the corresponding co-tensor
and finally we compute the $S_q^{\rm RIU}$ by performing the random walk
procedure described above. As we pointed before, we are mainly interested
in the cases $q=1$, $q=2$ and $q=100$. The latter value serves as an
approximation for the limining case $S_\infty ^{\rm RIU}$. The probability distributions of 
estimations of the minimal RIU entropy for random states are presented 
in Fig. \ref{sqst3qr}. 
The mean, second moment and standard deviation for these distributions are
reported in Table~\ref{tsq3q}. In the case of $q=1$ some analytical results are
available \cite{Slo98,Mir98}
\begin{equation}\label{sea}
\begin{array}{ll}
 \langle S_1 \rangle = \Psi(N+1) - \Psi(2) = \displaystyle \sum_{k=2}^N \frac1k,\\[3ex]
(\Delta S_1)^2 = \frac1{N+1} \left[ 2 \Psi^\prime(2) - (N+1) \Psi^\prime(N+1) \right],
\end{array}
\end{equation}
where $N$ is the size of the system and $\Psi(x)$ denotes the digamma function and
$\Psi^\prime(x)$ its derivative. In our case we set $N=8$ to obtain $\langle S_1
\rangle =1.718 $ and $\Delta S_1=0.160$, so
our numeric calculations are in good agreement with these analytical predictions.
We compute the maximum overlap of the state $\vert\psi\rangle$ with the closest
separable state by performing a random walk in the space of unitary matrices,
this quantity is denoted as $\lambda_{\rm max}^{\rm LU}$. The PARAFAC
decomposition of $\vert \psi \rangle$ yields a bound $\lambda_{\rm max} ^{\rm
PARAFAC}$ for this overlap.
A comparison between these three distributions is presented in Fig.~\ref{csqst3qr}.
\begin{table}[h]
\centering
\begin{tabular}{ | c | c c c| c c c| c c c|}
\hline
& \multicolumn{3}{c |}{$S_q(\vert \psi_{\rm typical} \rangle)$} & \multicolumn{3}{c| }{$S_q(\vert \psi_{\rm typical} \rangle_{\rm HOSVD})$} & \multicolumn{3}{c| }{$S_q^{\rm RIU}(\vert \psi_{\rm typical} \rangle)$}\\ \cline{2-10}
$q$ & $\langle S_q \rangle$ & $\langle S_q ^ 2 \rangle$ & $\Delta S_q$ & $\langle S_q \rangle$ & $\langle S_q ^ 2 \rangle$ & $\Delta S_q$ & $\langle S_q \rangle$ & $\langle S_q ^ 2 \rangle$ & $\Delta S_q$\\
\hline
1 & 1.717 & 2.973 & 0.161 & 1.132 & 1.407 & 0.355 & 0.907 & 0.858 & 0.184 \\[1.5ex]
2 & 1.534 & 2.397 & 0.224 & 0.812 & 0.794 & 0.368 & 0.650 & 0.461 & 0.197\\[1.5ex]
100 & 1.125 & 1.334 & 0.260 & 0.488 & 0.317 & 0.280 & 0.383 & 0.169 & 0.144\\
\hline
\end{tabular}
\caption{Mean value, second moment and standard deviation of the Renyi entropy of the
probability vector corresponding to a three-qubit random pure state (left),
 its corresponding co-tensor (center) 
 and the minimal RIU entropy for $q=1,2,100$.}
\label{tsq3q}
\end{table}

%
%
%

For any value of the R\'enyi parameter $q\ge 0$ one can ask for a state $\vert
\Phi_{q}^{\rm max}\rangle$, for which its minimal entropy $S^{\rm RIU}_{q}$
achieves its maximum value. It is known that in the case of three qubits the
maximal entangled state with respect the geometric measure is the  state $|W\rangle$
\cite{Che10, Aul10}, and hence the largest value of the RIU entropy reads
$S_\infty^{\rm RIU}(W) = - \log \lambda_{\rm max} (W) = - \log 4/9\approx
0.811$. We found numerically states for which $S_1^{\rm RIU}
(\Phi_{1}^{\rm max})= 1.277$, $S_2^{\rm RIU} (\Phi_{2}^{\rm
max})= 1.108$
attain the maximum values.
For $q=1,2$ the coefficient vectors in the basis
(\ref{acin1}) of these maximal states read
\begin{equation}\label{maxst3}
\begin{array}{ll}
\vert \Phi_1^{\rm max}\rangle = 0.27 \vert 000\rangle+0.377\vert 100\rangle+0.326\vert 010\rangle\\[1.5ex]
\hspace*{3cm}+0.363\vert 001\rangle+0.740 e^{-0.79 \pi i}\vert 111\rangle,\\[2ex]
\vert \Phi_2^{\rm max}\rangle =0.438 \vert 000\rangle+0.29\vert 100\rangle+0.371\vert 010\rangle\\[1.5ex]
\hspace*{3.2cm}+ 0.316\vert 001\rangle+0.698 e^{-0.826 \pi i}\vert 111\rangle.
\end{array}
\end{equation}
Furthermore, we  compute also $S_q^{\rm RIU}(\Phi_q^{\rm max})$ 
for other values of the R\'enyi parameter $q$ -- see Fig. ~\ref{sqst}.

\subsection{Distribution of $3$--tangle}\label{ss3tan}

The residual entanglement or 3-tangle $\tau$ was introduced by Coffman et al. \cite{Cof00}.
It quantifies the genuine entanglement of a system of three 
qubits $A$, $B$ and $C$ in the following sense. Let ${\cal C}_{A,B}$ and ${\cal
C}_{A,C}$ denote the concurrences~\cite{Hil97} of the density matrices of the
pairs of qubits $A,B$ and $A,C$, respectively. Since the qubits $BC$ can be
regarded as a single subsystem, we can ask for the entanglement between $A$ and
$BC$. Such a quantity will be denoted as ${\cal C}_{A,BC}$ and it is equal to $2
\sqrt{\det \rho_A}$ \cite{Cof00}, where $\rho_A = \mathrm{tr}_{BC} \rho_{ABC}$
is the reduced density matrix of the qubit $A$ when the partial trace respect to
$B$ and $C$ has been performed. The following inequality holds
\begin{equation}
 {\cal C}_{A,B}^2 + {\cal C}_{A,C}^2 \le {\cal C}_{A,BC}^2,
\end{equation}
and using it we define the \emph{3-tangle} as $\tau = {\cal C}_{A,BC}^2-{\cal
C}_{A,B}^2 - {\cal C}_{A,C}^2$. This quantity measures the amount of
entanglement between the qubit $A$ and the subsystem $BC$ that is not 
related to the entanglement in the pairs $A,B$ and $A,C$. The 3-tangle
$\tau$ is invariant under the permutation of  sub-systems and vanishes on all
states that are separable under any cut. The 3-tangle for the state (1) with
$n=3$ and $d=2$ is given by
\begin{equation}\label{tangle}
\tau(\vert \psi \rangle) = 4 \left \vert {\rm Det}_3 (C)\right\vert,
\end{equation}
where
\begin{equation}\label{det3}
 {\rm Det}_3 (C) = \sum \epsilon_{i_1,j_1} \epsilon_{i_2,j_2}\epsilon_{k_1,\ell_1}\epsilon_{k_2,\ell_2}\epsilon_{i_3,k_3}\epsilon_{j_3,\ell_3} C_{i_1,i_2,i_3}C_{j_1,j_2,j_3}C_{k_1,k_2,k_3}C_{\ell_1,\ell_2,\ell_3},
\end{equation}
is the Cayley hyperdeterminant of the tensor $C$.
Here $\epsilon_{i_1,i_2}$ stands for the Levi-Civita tensor of rank 2
and the sum is performed over all indexes. The 3--tangle is bounded by 0 and 1. 
For the $\vert GHZ \rangle$ state the residual entanglement attains its maximum value. 
Indeed, the pairwise concurrences ${\cal C}_{A,B}$ and ${\cal C}_{A,C}$ vanish and $\tau
= {\cal C}_{A,BC} = 1$, so this state is referred as a genuinely three-partite entangled state.

The hyperdeterminant ${\rm Det}_3 (C)$ is a homogeneous polynomial function
of degree 4,  first introduced by Caley \cite{Cay45}.
 It is invariant under the action of the group $SL(2,\mathbb{C})^{\otimes 3}$  \cite{Gel94}.
 This notion of
invariance plays an important role in the construction of entanglement monotones
for pure states of multiqubit systems \cite{Ost05}. Moreover, 
a classification of multipartite entangled states has been accomplished by analyzing
singularities of the hyperdeterminant~\cite{Miy03}.

\begin{figure}[htbp]
\centering
 \includegraphics[scale=0.63]{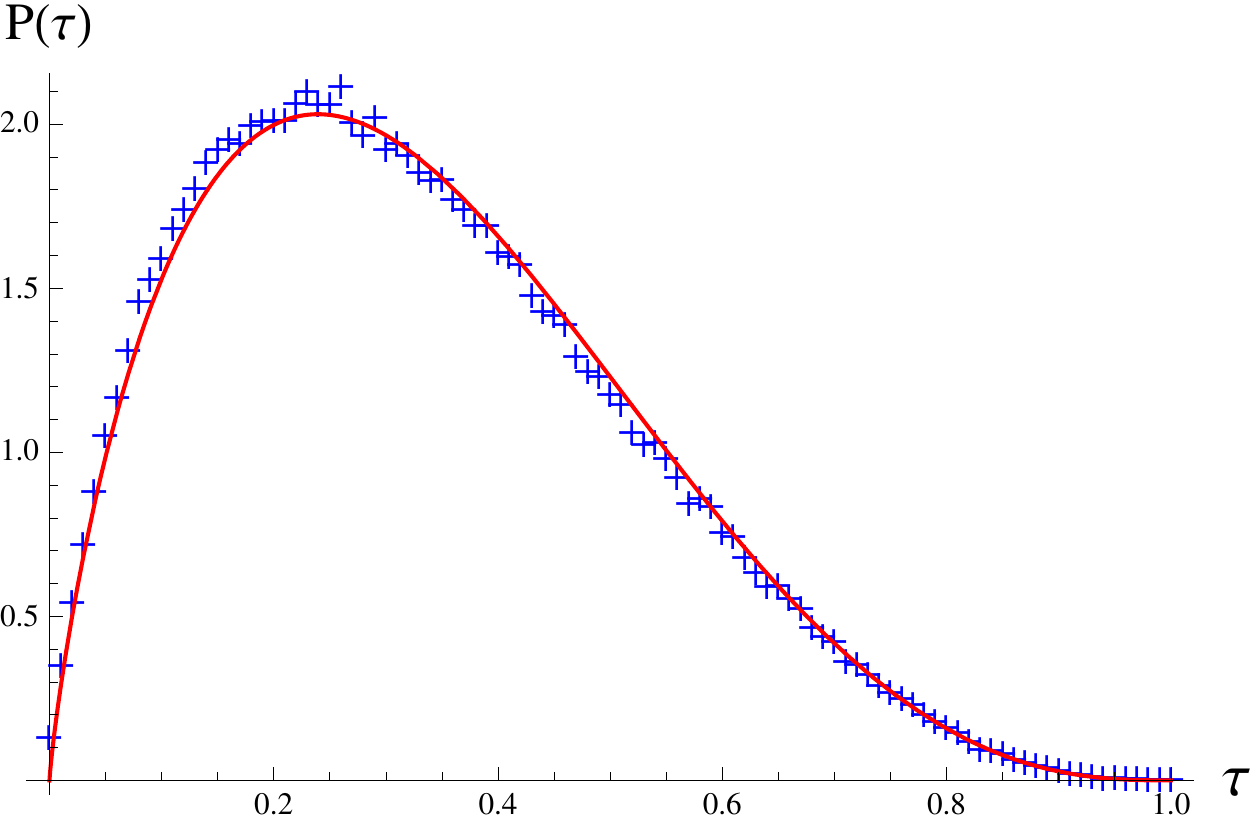}
\quad
 \includegraphics[scale=0.63]{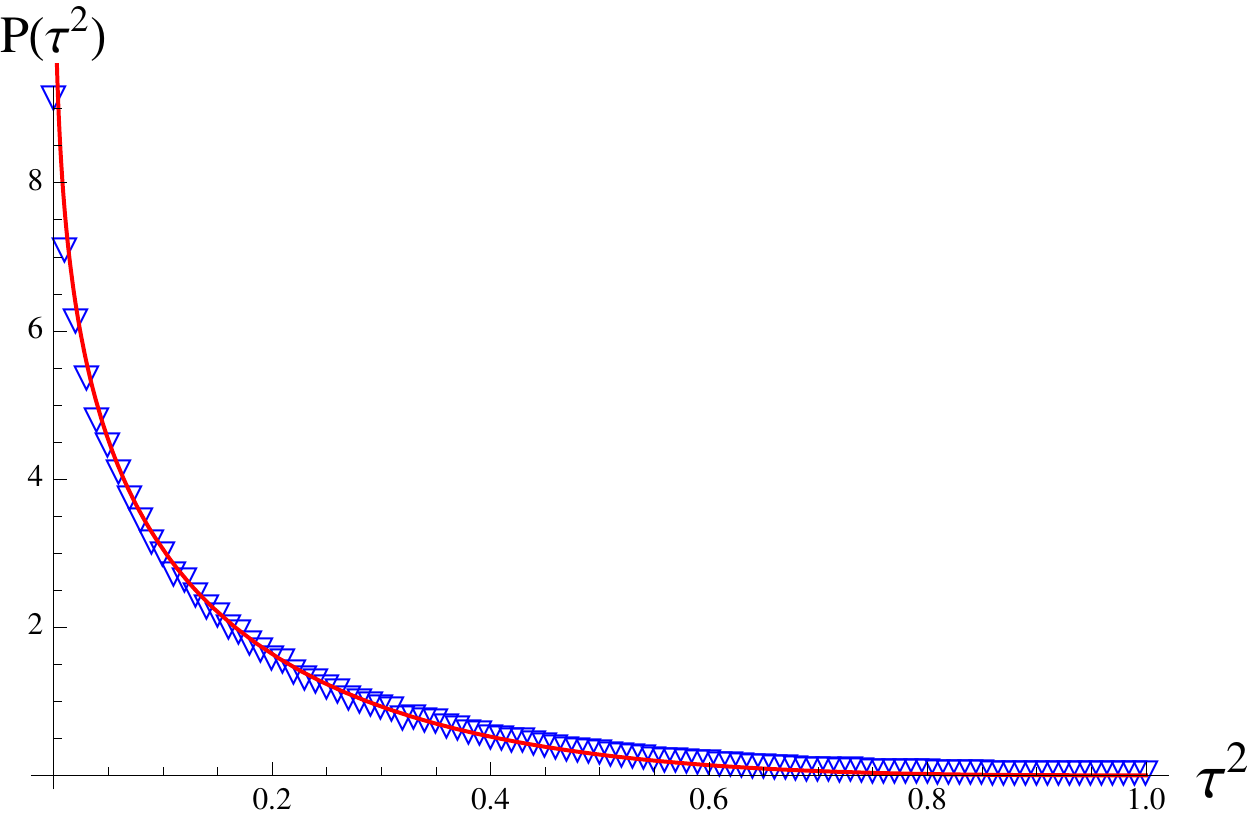}
\caption{(Color online) Distributions of the 3--tangle ({\color{blue} +}) and its square (${\color{blue} \triangledown}$) for three-qubit pure random states. The solid lines are approximations with Beta distributions (\ref{dtau}) at left, and (\ref{dtau2}) at right.}
\label{dtauim}
\end{figure}

Now we will discuss properties of typical quantum states of a three--qubit system.
Kendon et al. ~\cite{Ken02} studied the distribution of 
3-tangle for an ensemble
of random pure states drawn according to the Haar measure.
By direct integration of equation (\ref{tangle}) with
respect the unitary invariant measure
they obtained the average value, $\langle \tau \rangle = \frac13$.
In Fig.~\ref{dtauim} we show the
probability densities of $\tau$ and $\tau^2$ evaluated over a sample of $10^5$
random pure states. Moreover, the first six even moments of $\tau$ can be
calculated by symbolic integration using the Beta integral (see Appendix
\ref{sec:moments}). The distribution of 3-tangle can be approximated
by the Beta distribution $\mathrm{Beta}(\alpha,\beta; \tau)$,
using the moment method to estimate the parameters 
$\alpha = \langle \tau \rangle \left(\frac{\langle \tau \rangle (1- \langle \tau \rangle)}{\langle \tau^2 \rangle - \langle \tau \rangle^2} -1 \right) = \frac{31}{17}$ and 
$\beta  = \left(1 - \langle \tau \rangle\right) \left(\frac{\langle \tau \rangle (1- \langle \tau \rangle)}{\langle \tau^2 \rangle - \langle \tau \rangle^2} -1 \right) = \frac{62}{17}$.
Hence we get
\begin{equation}\label{dtau}
 P_B(\tau) = \mbox{Beta}\left(\frac{31}{17},\frac{62}{17};\tau\right).
\end{equation}
By virtue of the chain rule we get an approximated distribution function for $\tau^2$
\begin{equation}\label{dtau2}
 P_B(\tau^2) = \frac1{2\sqrt \tau} \mbox{Beta}\left(\frac{31}{17},\frac{62}{17};\sqrt{\tau}\right).
\end{equation}
The above distributions are presented in Fig.~\ref{dtauim} together with 
the estimation obtained by numerical simulations. In order to
show the accuracy of our approximation, we show in Table~\ref{dtauta} a
comparison between the first 6 even moments of the distribution (\ref{dtau}) and the
moments of $P(\tau)$ obtained by symbolic integration.
\begin{table}[htbp]
\centering
\begin{tabular}{| l | l | l |}
\hline
$k$ &$k$-th moment of $P_B(\tau)$ & $\langle \tau^k\rangle$,  $k$-th moment of $P(\tau)$ \\[1ex]
\hline
1 & 1/3 & 1/3 \\[1ex]

2 & 8/55 & 8/55\\[1ex]
 
4 & 533/12573 $\approx$ 0.04239 &128/3003 $\approx$ 0.04262 \\[1ex]
 
6 & 30914/1819783 $\approx$ 0.01699 & 7168/415701 $\approx$ 0.01724 \\[1ex]
 
8 & 112955/13778357 $\approx$ 0.008198 & 98304/11685817 $\approx$ 0.008412 \\[1ex]
 
10 & 1840340/411553533 $\approx$ 0.004472 & 262144/56497545 $\approx 0.00464$ \\[1ex]

12 & 672000151/252556684751 $\approx$ 0.002661 & 4194304/1502700975 $\approx$0.002791 \\[1ex]
\hline
\end{tabular}
\caption{Comparison of the first six even moments of $\tau$ approximated with the equation (\ref{dtau}) (center column) and computed using the symbolic integration package (right column).}
\label{dtauta}
\end{table}
%

\section{Four qubits}\label{sfouqub}

In this section we compute both the minimal RIU entropy and the hyperdeterminant
for several exemplary 4-qubit states, we also analyze some typical properties of
these quantities.
\subsection{Minimal decomposition entropy}

According to Carteret {\it et al} \cite{Car00}, a given state $\vert \Phi \rangle \in
{\cal H}_2^{\otimes 4}={\cal H}_{16}$ can be written using twelve terms 
\begin{equation}\label{4qdecom}
 \vert \Phi \rangle = \sum_{i,j,k,\ell} c_{i,j,k,\ell} \vert i,j,k,\ell \rangle, ~ \mbox{with } c_{0,1,1,1}=c_{1,0,1,1}=c_{1,1,0,1}=c_{1,1,1,0}=0,
\end{equation}
and the other coefficients $c_{i,j,k,\ell}$ are real and non-negative if at most one of the indexes differs from 1. The state $|GHZ_4\rangle$
 is obtained by choosing $c_{0,0,0,0}=c_{1,1,1,1}=1/\sqrt 2$. Hence, $S_0^{\rm RIU}(GHZ_4)=2$. Moreover, we define a
 state $\vert A_{12}\rangle $ for which the non-vanishing coefficients 
in the above expansion are $c_{i,j,k,\ell} = 1/\sqrt{12}$. 
In this case $S_0^{\rm RIU} (A_{12})=12$.

\begin{figure}[htbp]
\centering
 \includegraphics[scale=0.75]{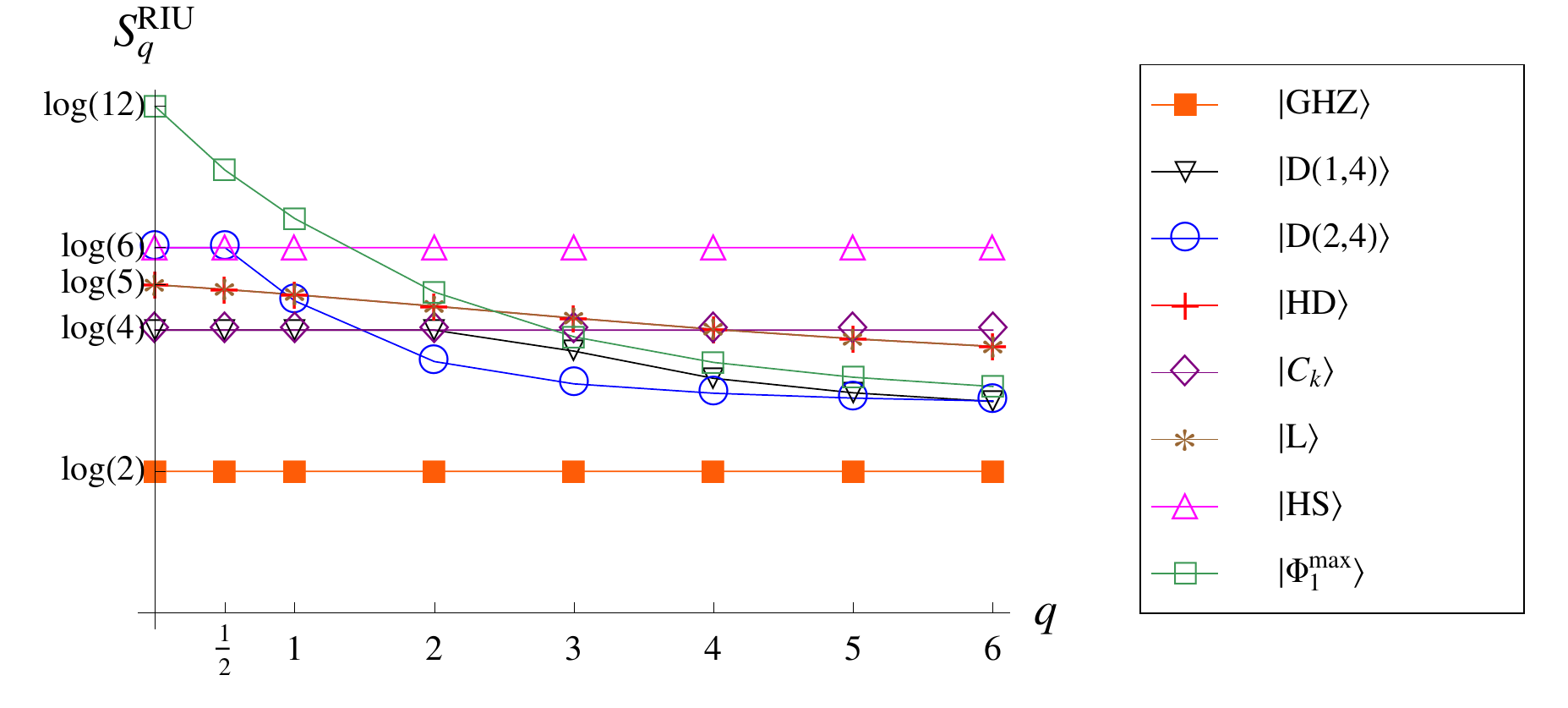}
 \caption{(Color online) The minimal RIU entropy $S_q^{\rm RIU}$ computed for several four qubit states as function of the parameter $q$. }
 \label{sqst4}
\end{figure}

A proper basis for $n$ qubit permutation invariant states is constituted by 
the so-called Dicke states
\begin{equation}
 \vert D(n,k) \rangle = \binom n k^{-1/2} \sum_{\rm permutations} \vert \underbrace{1\cdots1}_{k} \underbrace{0\cdots0}_{n-k} \rangle.
\end{equation}
Note that $\vert W \rangle = \vert D(n,1) \rangle$, $\vert {\widetilde W}\rangle
= \vert D(n,n-1) \rangle$ and $\vert GHZ_n \rangle = (\vert D(n,0) \rangle + \vert
D(n,n) \rangle)/\sqrt 2$. In computations below we will consider the case of
$n=4$.
The calculation of the minimal RIU entropy for permutation invariant states can be turned into a one-variable optimization by taking $U_{loc} = U(p)^{\otimes 4}$. 
We found $S_q^{\rm RIU}(GHZ_4)=\log 2$, for all $q$ and $S_1^{\rm RIU}(D(4,1))=\log 4$; for the other values of $q$ minimization will be accomplished numerically. On the other hand, the optimal decomposition of the state $\vert D(4,2)\rangle$  is obtained by taking $U_{loc} = U(1/2)^{\otimes 4}$. Accordingly, it is
possible to get a compact formula for its minimal RIU entropy
\begin{equation}\label{d24}
 S_q^{\rm RIU} (D(4,2)) = \frac 1{1-q} \log [2^{1-3 q} 3^{-q} (3+3^{2q})],
\end{equation}
hence we get $S_1^{\rm RIU}(D(4,3))= \log (8/\sqrt 3)$ by taking the
limit $q\rightarrow 1$. On the other hand, when $q\rightarrow \infty$ one
arrives to $S_\infty ^{\rm RIU}(D(4,2)) = - \log (3/ 8)$, which is consistent
with the well known result of the overlap of $\vert D(4,2) \rangle$ with the
closest separable state \cite{Hub09}. Note that by symmetry $S_q^{\rm
RIU}(D(4,3))=S_q^{\rm RIU}(D(4,1))$. In Fig.~\ref{sqst4} we compare the results
obtained in this way with those acquired by performing a random walk over the space
of unitary matrices for several values of $q$.

Consider also some other exemplary  4-qubit states. The 
hyperdeterminant state 
\begin{equation}\label{hdst}
\vert {\rm HD}\rangle = \frac1{\sqrt{6}} \left(\vert 1000\rangle + \vert 0100\rangle+\vert0010\rangle +\vert 0001\rangle+\sqrt2\vert 1111\rangle\right),
\end{equation}
maximizes the 4-qubit hyperdeterminant \cite{Als13}. The minimal RIU entropy for the former state can be computed in the same way as for the Dicke state (\ref{d24}), as it is shown \cite{Hub09,Hay09,Wei10} 
that the state (\ref{hdst}) presents the optimal decomposition minimizing the RIU entropy for any $q$. 
Then we obtain
\begin{equation}
 S_1^{\rm RIU}\left ({\rm HD}\right) = \frac43 \log 3, \quad S_q^{\rm RIU}\left ({\rm HD}\right) = \frac{1}{1-q} \log\left(\frac{6^q}{4+4^q}\right), \mbox{ for } q\neq 1.
\end{equation}
In the limit $q\rightarrow\infty$ we get $S_\infty^{\rm RIU}\left ({\rm
HS}\right) = -\log(2/3)$ which is consistent with the maximum overlap with the
closest separable states computed according to the Ref. \cite{Hub09}.
On the other hand, the {\it cluster states} identified by Gour and Wallach
\cite{Gou10}
\begin{equation}
\begin{array}{ll}
 \vert {\rm C}_1\rangle = \frac12\left(\vert 0000\rangle+\vert0011\rangle+\vert1100\rangle-\vert1111\rangle\right),\\[1.1ex]
 \vert {\rm C}_2\rangle = \frac12\left(\vert 0000\rangle+|0110\rangle+|1001\rangle-\vert1111\rangle\right),\\[1.1ex]
 \vert {\rm C}_3\rangle = \frac12\left(\vert 0000\rangle+|0101\rangle+|1010\rangle-\vert1111\rangle\right),
\end{array}
\end{equation}
are the  4-qubit states that maximize the R\'enyi $\alpha$-entropy of
partial trace for $\alpha\ge 2$. They also found that the state
\begin{equation}
\begin{array}{ll}
 \vert {\rm L}\rangle = \frac1{\sqrt{12}} 
\bigl[ (1+w)(\vert 0000\rangle+\vert 1111\rangle)+(1-w)(\vert 0011\rangle+\vert1100\rangle)\\[2ex]
 \hspace*{2cm}+w^2(\vert 0101\rangle + \vert 1001\rangle +\vert 1010\rangle)\bigr] 
,\quad \mbox{with } w= \exp(\frac{2i\pi}{3}),
 \end{array}
\end{equation}
 maximizes the average Tsallis $\alpha$-entropy 
of the partial trace for all $\alpha>2$. The minimal RIU entropy of the former states is
shown in Fig.~\ref{sqst4} for several values of $q$. 
Based on our numerical calculations for the cluster
states we conjecture that $S_q^{\rm RIU} ({\rm C}_k) = \log 4$ with $k=1,2,3$ for any
value of the R\'enyi parameter.

\begin{figure}[htbp]
\centering
 \includegraphics[scale=0.64]{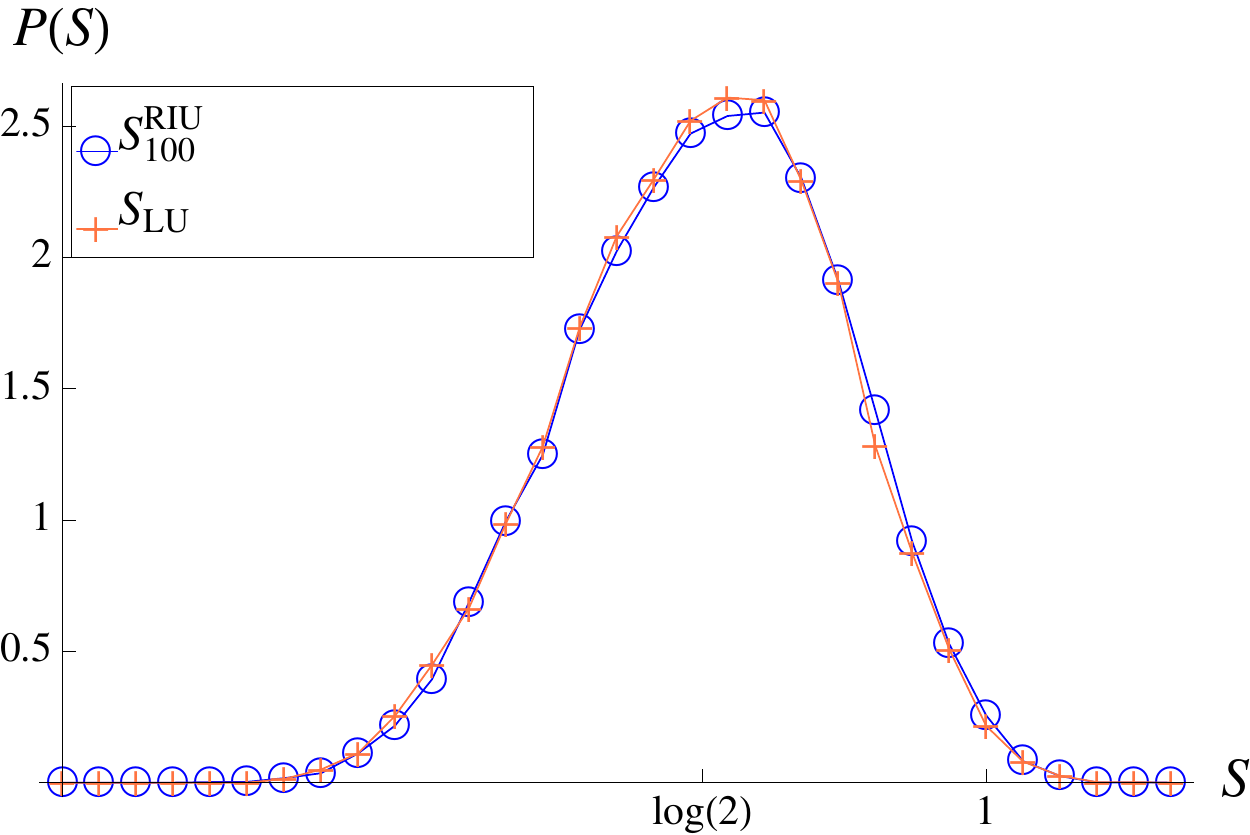}
 \caption{(Color online) Comparison between the distributions of $S_{100}^{\rm RIU}$ ({$\color{blue} \circ$}) and $S^{\rm LU}=-\log \lambda_{\rm max}^{\rm LU}$ ($+$) for a four--qubit system.} 
\label{csqst4qr}
\end{figure}

We also consider the 4-qubit state found by Higuchi and Sudbery \cite{Hig00}
\begin{equation}
 \vert{\rm HS}\rangle = \frac16 \bigl[ \vert0011\rangle+\vert1100 \rangle
 +w(\vert0101\rangle+\vert1010\rangle)+w^2(\vert0110\rangle+\vert1001\rangle ) \bigr],
\label{HSstate}
\end{equation}
where $w=\exp(2i\pi/3)$, which has maximum average von Neumann entropy
of partial traces averaged over all possible 3 splittings of 4 qubit system
into two bipartite systems. Numerical values of $S_q^{\rm RIU}({\rm HS})$ are
shown in Fig.~\ref{sqst4}.
We are also interested in the typical properties of the minimum of $S_q^{\rm
RIU}$ for four qubits. For an ensemble of $10^5$ random pure states we compute
$S_q(\vert \psi \rangle)$, $S_q(\vert \psi \rangle_{\rm HOSVD})$ and $S_q^{\rm
RIU} (\vert \psi \rangle)$ taking $q=1,2,100$. The corresponding distributions
are shown in Fig.~\ref{sqst4} and in Table~\ref{sqst4t}. In the first case and
with $q=1$ we can compare our results with those provided by the analytic expression 
(\ref{sea}), $\langle S_1 \rangle = 2.381$ and $\Delta S_1 = 0.124$. To provide an
approximation of $S_\infty^{\rm RIU}$ we analyzed the entropy with the Renyi parameter
$q=100$. In Fig.~\ref{csqst4qr} we show the distributions of the minimal RIU entropy $S_{100}^{\rm RIU}$ and $S_{\rm
max}=-\log \lambda_{\rm LU}$, where $\lambda_{\rm max}^{\rm LU}$ is the maximum
overlap with the closest separable state computed through a random walk in the
space of unitary matrices.
\begin{table}[htbp]
\centering
\begin{tabular}{| c| c c c| c c c| c c c|}
\hline
& \multicolumn{3}{c |}{$S_q(\vert \psi_{\rm typical} \rangle)$} & \multicolumn{3}{c| }{$S_q(\vert \psi_{\rm typical} \rangle_{\rm HOSVD})$} & \multicolumn{3}{c| }{$S_q^{\rm RIU}(\vert \psi_{\rm typical} \rangle)$}\\ \cline{2-10}
$q$ & $\langle S_q \rangle$ & $\langle S_q ^ 2 \rangle$ & $\Delta S_q$ & $\langle S_q \rangle$ & $\langle S_q ^ 2 \rangle$ & $\Delta S_q$ & $\langle S_q \rangle$ & $\langle S_q ^ 2 \rangle$ & $\Delta S_q$\\
\hline
1 & 2.381 & 5.686 & 0.124& 2.038 & 4.234 & 0.283 & 1.633 & 2.687 & 0.145 \\[1.5ex]
2 & 2.159& 4.698 & 0.190 & 1.601 & 2.727 & 0.403 & 1.199 & 1.473 & 0.192\\[1.5ex]
100 & 1.60 & 2.635 & 0.254 & 1.027 & 1.219 & 0.405 & 0.701 & 0.513 & 0.145\\
\hline
\end{tabular}
\caption{Mean value, second moment and standard deviation of the R\'enyi entropy of a four-qubit random pure state (left), its corresponding co-tensor (center) and the minimal of the RIU entropy for $q=1,2,100$.}
\label{sqst4t}
\end{table}
%

%
%
\begin{figure}[tbp]
 \centering
 \begin{tabular}{ccc}
 \includegraphics[scale=0.52]{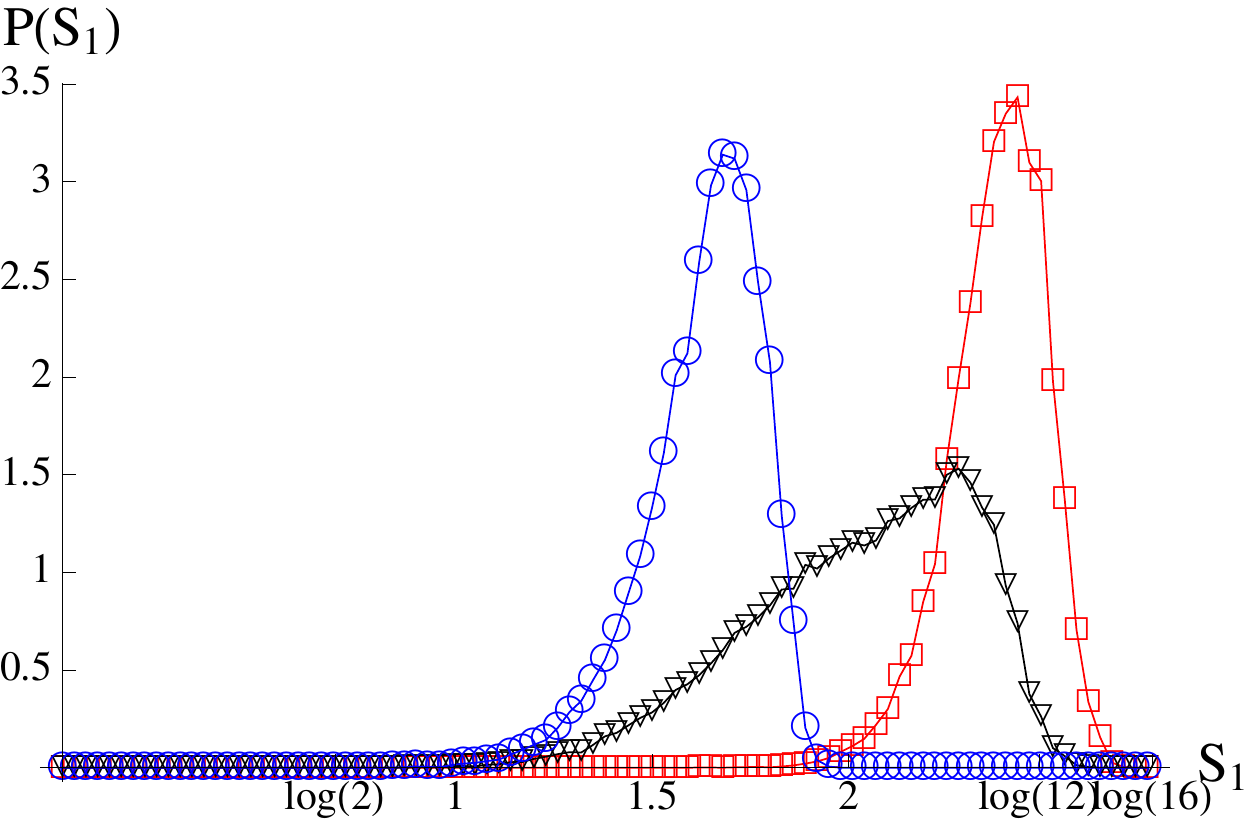}& \hspace*{0.5cm}& \includegraphics[scale=0.5]{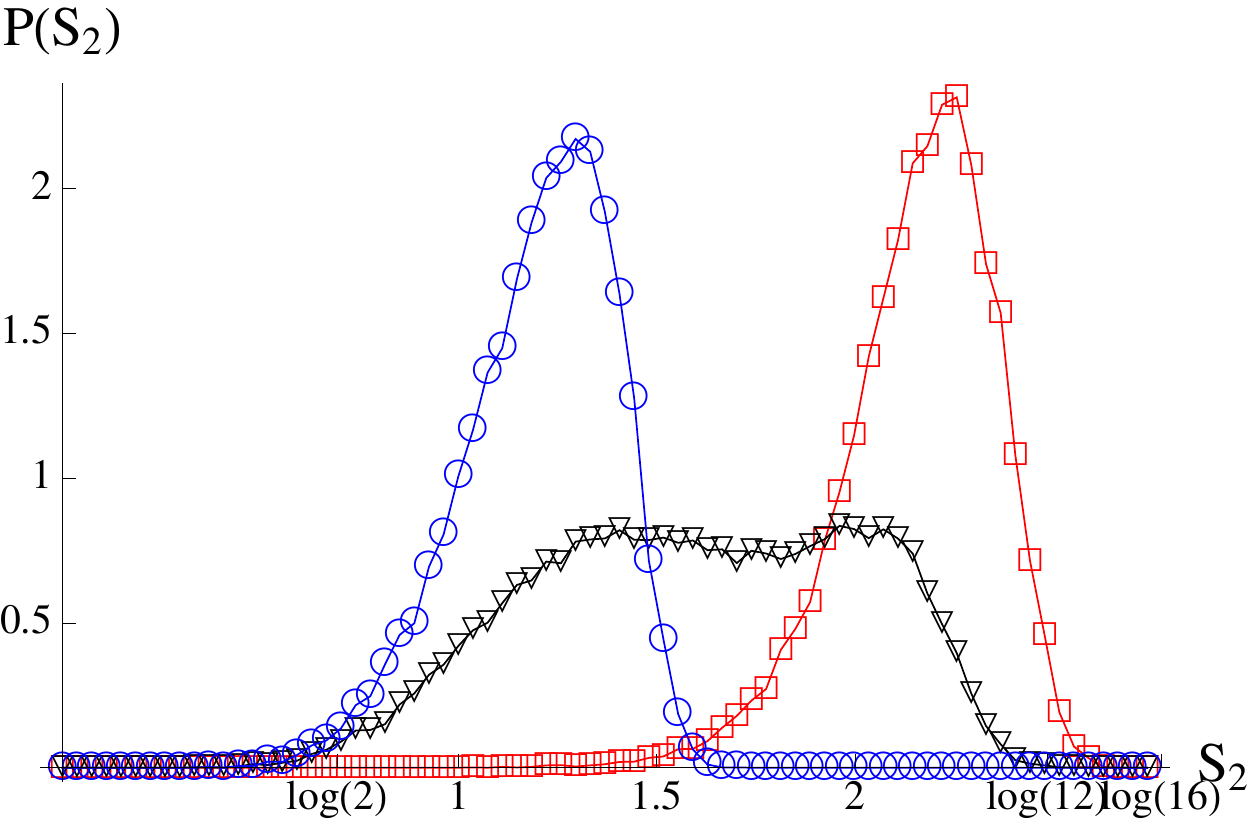} \\[1ex]
 a) & & b) \\[1ex]
\includegraphics[scale=0.55]{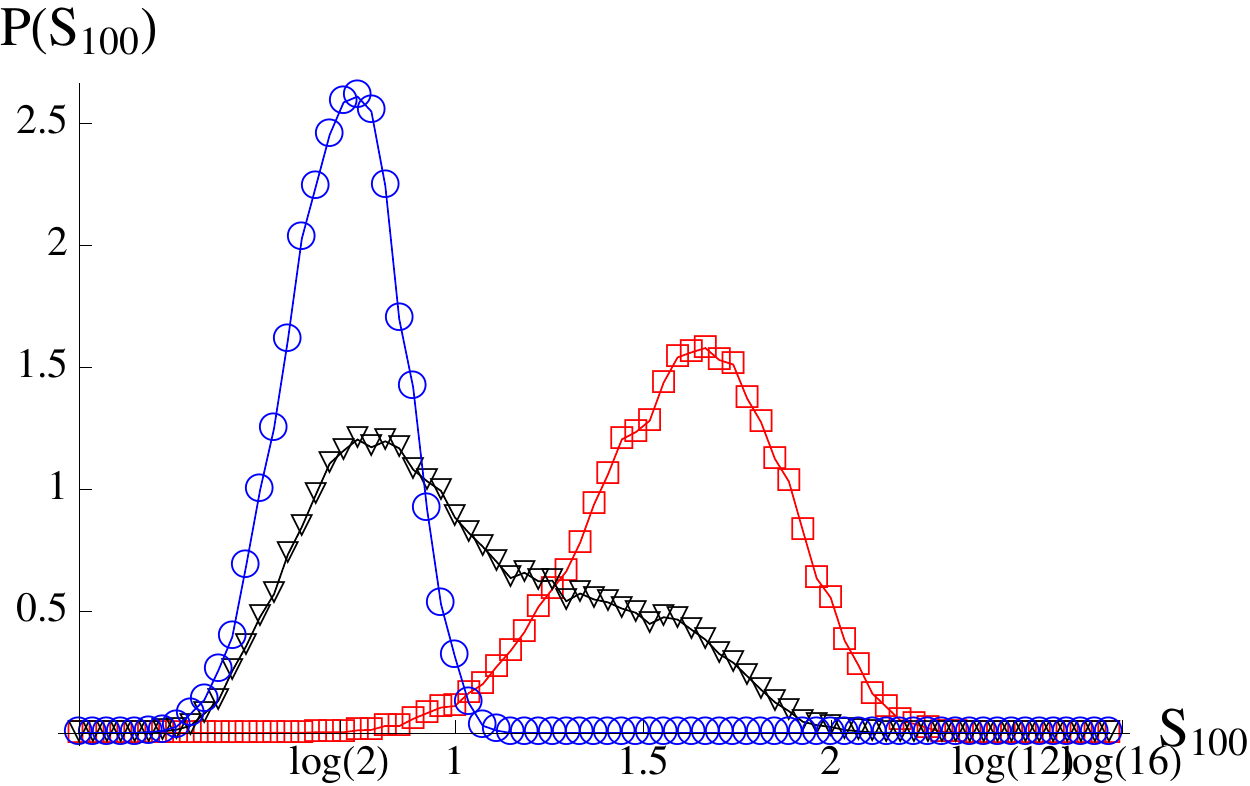} & & \includegraphics[scale=0.51]{labelsq.pdf}\\[1ex]
 c) & {} & {}
 \end{tabular}
\caption{(Color online) Distributions of the R\'enyi entropy of a four--qubit random state probability vector ({$\color{red} \square$}), the R\'enyi entropy of the corresponding co-tensor ($\triangledown$) and the minimal RIU entropy ({$\color{blue} \circ$}) for (a) $q=1$, (b) $q=2$ and (c) $q=100$.}
\label{sqst4qr}
\end{figure}

%

As in the three qubit case, we are looking for the maximal states of the minimal
RIU entropy. For the maximal symmetric state with respect to the geometric
measure of entanglement $\vert \Phi_4\rangle = \sqrt{1/3}\vert D(4,0)\rangle +
\sqrt{2/3}\vert D(4,3)\rangle$ the RIU entropy yields $S_\infty^{\rm RIU}
(\Phi_4) = - \log(1/3)\approx  1.099$ \cite{Aul10}. 
Moreover, our numerical results allow us to conjecture that
$S_{\infty}^{\rm RIU}({\rm HS}) = -\log(2/9)\approx 1.504$, 
is the largest value of the minimal RIU entropy with $q=\infty$ among the four
qubit states.

For $q=1$ we found numerically a state  $\vert \Psi_1^{\max} \rangle$
for which $S_1^{\rm RIU}(\Psi_1^{\rm max})=1.934$ is maximal. In the decomposition (\ref{4qdecom}) this state reads  
\begin{equation}\label{maxst4}
\begin{array}{ll}
\vert \Phi_1^{\rm max}\rangle = 0.630 \vert 0000\rangle+0.281\vert 1100 \rangle+0.202\vert 1010\rangle+0.24\vert 0110\rangle\\[1.5ex]
\hspace*{1.5cm}+0.232 e^{0.494\pi i} \vert 1110\rangle+0.059\vert1001\rangle+0.282\vert 0101\rangle\\[1.5ex]
\hspace*{2cm}+0.346e^{-0.362\pi i}\vert1101\rangle+0.304\vert 0011\rangle+0.218e^{0.626\pi i}\vert1011\rangle\\[1.5ex]
\hspace*{2.3cm}+0.054 e^{-0.725\pi i} \vert 0111\rangle+0.164 e^{0.372\pi i}\vert 1111\rangle.
\end{array}
\end{equation}
Further numerical tests support the conjecture that the state $|{\rm HS}\rangle$,  for which  $S_2^{\rm RIU} = \log 6$,
maximizes the minimal RIU entropy for $q\ge2$.
\subsection{Distribution of the hyperdeterminant $\vert {\rm Det}_4\vert $}

In the case of 4-qubit states the hyperdeterminant ${\rm Det}_4$ is a polynomial of degree 24. It can be constructed following the Schl\"afli's procedure \cite{Miy03}. 
In analogy to the three-tangle we consider the following function of the coefficients tensor $C$ of the state $\vert \psi\rangle\in {\cal H}_{16}$,
\begin{equation}\label{det4t}
 T(\vert \psi\rangle) = 2^6 3^9 \vert {\rm Det}_4 (C) \vert,
\end{equation}
For a separable four--qubit state one has $T(\vert \psi_{\rm sep}\rangle ) = 0$
but it vanishes also for the states $\vert D(1,4)\rangle$, $\vert D(2,4)\rangle$, $\vert C_k\rangle$, $\vert HS\rangle$ and $\vert GHZ \rangle$.
 
Alsina and Latorre found recently \cite{Als13}
that both $\vert HD\rangle$ and $\vert L \rangle$ maximize the hyperdeterminant and they 
also discussed relation between these states for which $T(|HD\rangle)=T(|L\rangle)=1$.
Furthermore, numerical simulations indicate that they have the same minimal RIU entropy $S_q^{\rm RIU}$.

\begin{figure}[htbp]
 \centering
 \includegraphics[scale=0.7]{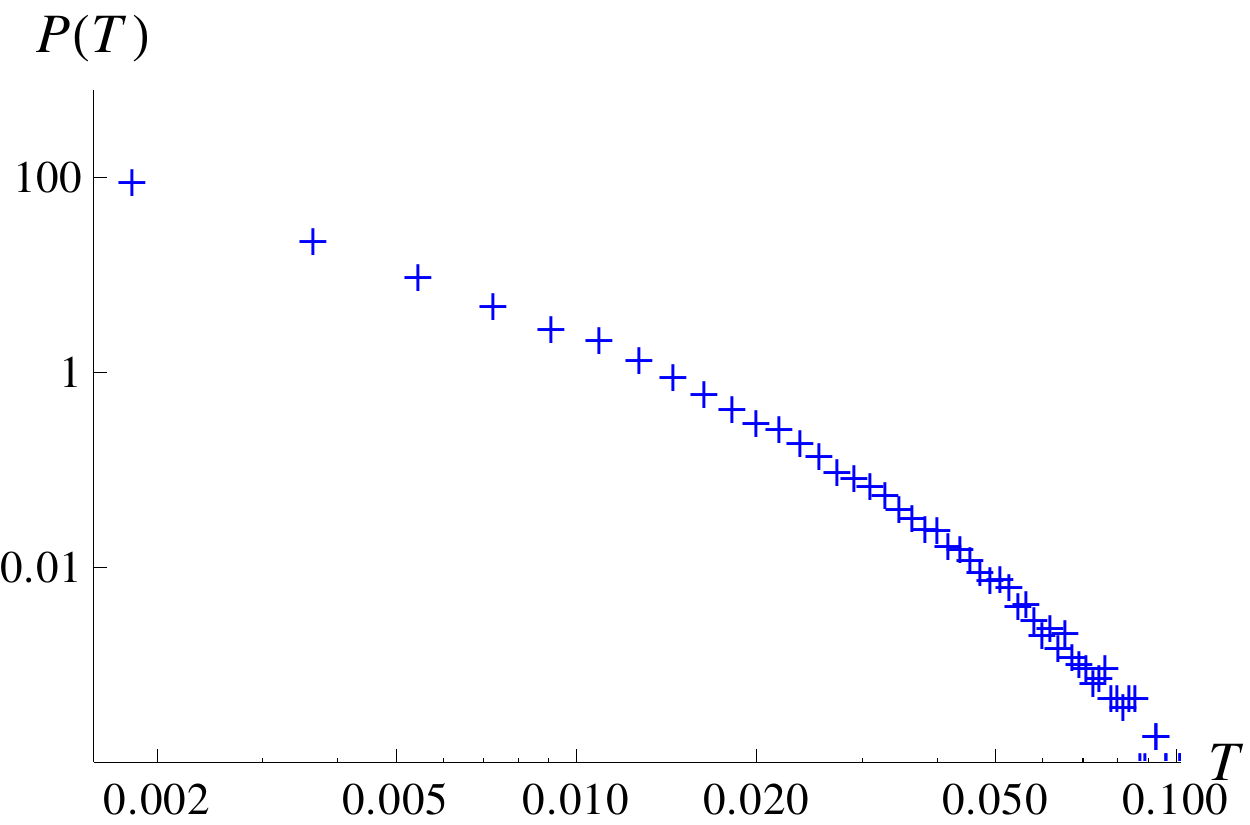}
\caption{(Color online) Distribution of absolute value of the hyperdeterminant
$P(T)$ ({\color{blue} +}) for four qubit pure random states in a log-log plot.}
\label{dt4dist}
\end{figure}

We evaluate the quantity (\ref{det4t}) over an ensemble of $10^7$ random pure states.
The mean and the standard deviation read 
 $\langle T \rangle = 9.74\times 10^{-4}$ and $\langle \Delta T \rangle = 2.39 \times 10^{-3}$, respectively, 
while the corresponding distribution is shown in Fig.~\ref{dt4dist}.

\section{Three qudits: Asymptotic case.}\label{sasyc}

Statistical properties of random tensors and their asymptotic limit
became a subject of an intensive research \cite{AHH10,Gura11}.
In this section we study random tenors of dimension three
and arbitrary size, which describe generic states of a system
composed of three qudits. 
We analyze the bounds for the geometric measure of entanglement 
of such states provided by the tensor decompositions HOSVD and PARAFAC.

\begin{figure}[htbp]
\centering
\includegraphics[scale=0.73]{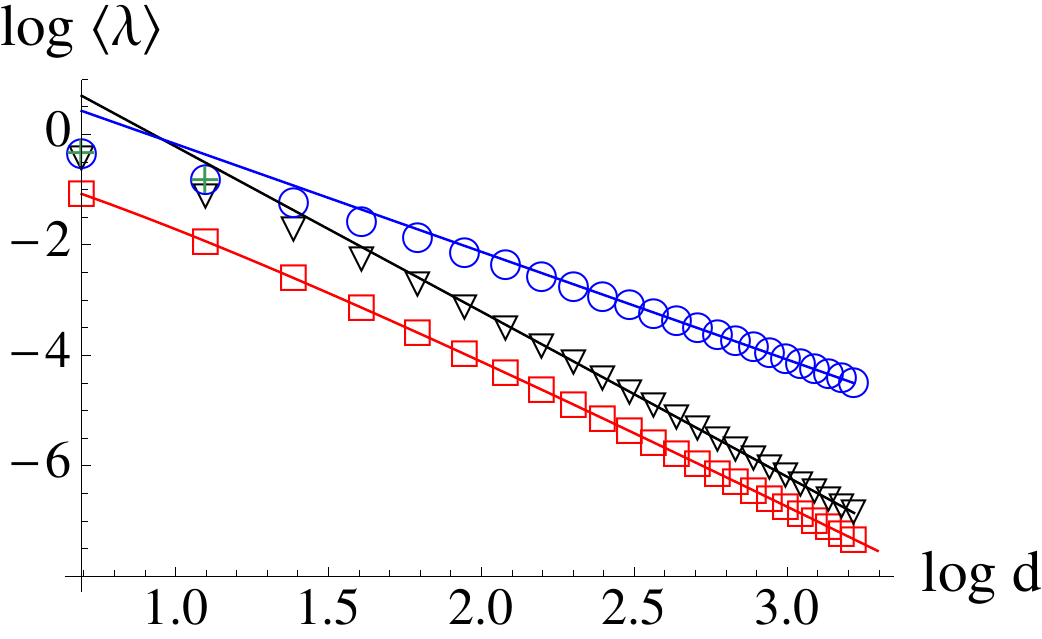}\quad\quad
\includegraphics[scale=0.73]{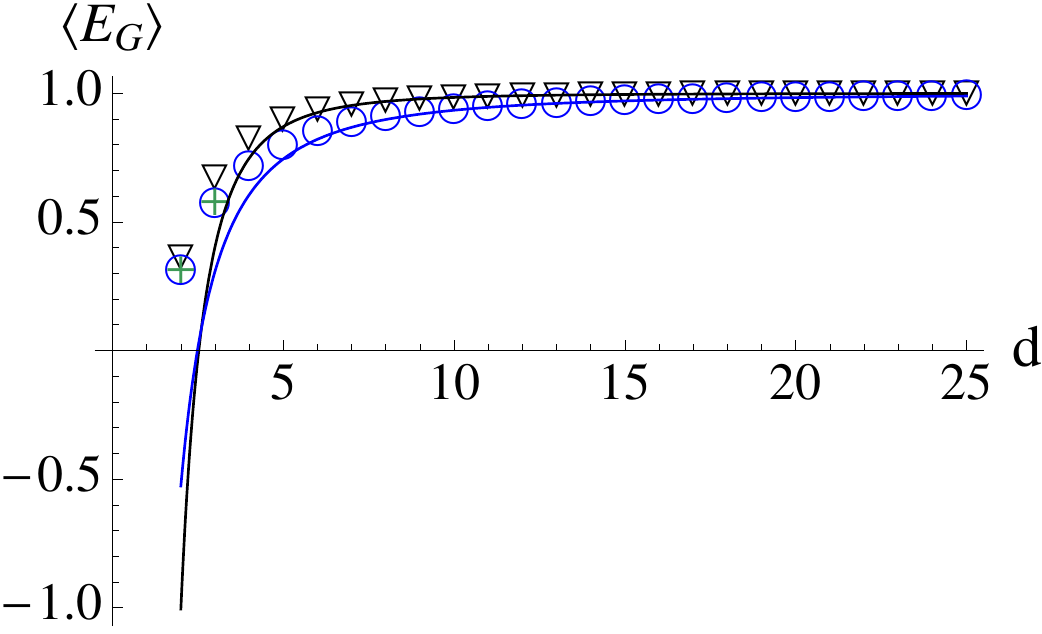}
\caption{(Color online)  The mean of the maximum component (left) and the geometric measure of entanglement (right) for random states of a tri--partite system
as a function of the qudit size. 
Bullets correspond to numeric simulations: ({\color{red} $\square$}) stands for the greatest tensor element $\lambda_{\max}$; ($\triangledown$) the maximum component of the HOSVD co-tensor $\lambda_H$; ({\color{blue} $\circ$}) stands for PARAFAC overlap $\lambda_P$ and ({\color{greenm} $+$}) refers to the overlap with the closest separable state maximized by LU. The solid red line ({\color{red} \bf ---}) is the result (\ref{maxelem}) with $N=d^3$, the solid lines ({\color{black} \bf ---}) and ({\color{blue} \bf ---}) are the best linear fits for HOSVD and PARAFAC, respectively}
\label{3quditsd}
\end{figure}

Consider a typical state $ \vert \psi \rangle \in {\cal H}_d ^{\otimes 3}$ drawn from the Haar measure. Let $\lambda_{\rm max}$ and $\lambda_{H}$ denote the largest component of the probability vectors $p(\vert \psi \rangle)$ and $p(\vert \psi \rangle_{\rm HOSVD})$, respectively, and consider the overlap $\lambda_P=\vert\langle \psi \vert \psi_P\rangle\vert^2$, where $\vert \psi_P\rangle$ is the state (\ref{tensor3}). In the cases $d=2,3$, we also evaluate the maximum overlap of $\vert \psi \rangle$ with the closest separable state by performing a random walk optimization
procedure in the space of unitary matrices. 
Fig.~\ref{3quditsd} (a) shows the average value of four quantities as function of the size of the quit $d$ computed over an ensemble of $10^5$ random states. 
The mean value of $\lambda_{\rm max}$ can be expressed \cite{ALPZ15}
in terms of the harmonic numbers $H_N$,
\begin{equation}\label{maxelem}
 \langle \lambda_{\max} \rangle = \frac1{N} \sum_{j=1}^{N} \frac1j =H_{N} /N,
\end{equation}
where $N=d^3$, in our case. 
Therefore, for a random pure state of three qudits, 
the average largest overlap to a pure state scales as $d^{-3}$.
Performing decompositions for an ensemble of such random tensors
we find that for large $d$ the largest overlaps 
optimized by HOSVD and PARAFAC behave as

\begin{equation}\label{overlbounds}
 	\langle \lambda_{\rm H} \rangle \sim  d^{-2.99},\qquad
	\langle \lambda_{\rm P} \rangle \sim  d^{-1.95}.
\end{equation}

In both cases, the corresponding bound for the geometric measure of entanglement reads $\langle E_G\rangle \sim 1-\langle \lambda_{\rm k} \rangle$ where $k=H,P$. Fig.~\ref{3quditsd} (b) shows the geometric measure of entanglement for a three--qudit system as function of the qudit size. Moreover, based on the scaling of $\langle \lambda_{\rm P}\rangle$ we conjecture that the maximum overlap with respect to the closest separable state for a three--qudit system scales as $\langle \lambda_{\rm max} \rangle \sim  d^{-2}$ for large $d$.

\section{Conclusions}

We have analyzed the R\'enyi--Ingarden--Urbanik entropy 
of pure states of quantum multipartite systems
minimized over all local unitary operations.
For separable states such a quantity is zero irrespectively of the value of the R\'enyi parameter $q>0$.
In general it is not easy to get analytical results for a given state
and arbitrary value of the R\'enyi parameter $q$. 
In the special case of permutation invariant states the problem becomes easier, as
minimization in the space of local unitary operators can be turned into an optimization of a one-variable function.

We computed the minimal RIU entropy of several representative three and four qubit states for various values of the R\'enyi parameter.
Some particular states, which maximize the minimal RIU entropy 
for  $q=1$ and $q=2$, were identified. 
Note that this quantity can be considered as a measure of pure states
entanglement in multipartite systems, and for $q\to\infty$
it becomes a function of the geometric measure of entanglement.
In the case of three qubits the latter quantity is maximal for the state $|W\rangle$,
while for four qubits, it achieves maximum for the state (\ref{HSstate}) of 
Higuchi and Sudbery \cite{Hig00}.

Furthermore, we analyzed the distribution of the minimal RIU entropy 
for an ensemble of random for some
selected values of the Renyi parameter, $q=1,2,100$. 
Our numerical simulations demonstrate usefullness of the PARAFAC algorithm
which allows one the find the principal component of a tensor
representing a multipartite pure state
and estimate its geometric measure of entanglement.

We studied also the distribution of $3$--tangle 
for random pure states of  three qubit system
and derived a few even moments of this distribution.
Numerical results show that this distribution
may be approximated by a Beta distribution. 
In the case of four qubits we analyzed the distribution 
of the absolute value of the hyperdeterminant of a random pure state.

Finally, the behavior of the maximum overlap of a random state of a system
consisting of three subsystems of size $d$ and 
the closest separable states was investigated in the asymptotic limit. 
Although the size of the largest component of a random state scales as $d^{-3}$, 
numerical results obtained by the PARAFAC decomposition of a tensor
allow us to conjecture that the minimal overlap scales 
in this case as $d^{-2}$. This is consistent with an analytic
result based on the Marchenko-Pastur asymptotic distribution of singular
values of random matrices, which gives an upper bound $d^{-1}$.

\section{Acknowledgments}

We are thankful to D. Alsina and J. I. Latorre for fruitful discussions and for sharing their results
prior to publication.
ME is thankful to the Jagiellonian University for kind hospitality and support during his stay in Cracow.
We acknowledge support of the Mexican National Council for Science and Technology (Conacyt) (M.E.)
and  by the grants number DEC-2011/02/A/ST1/00119 (K.\.Z{})
and DEC-2012/04/S/ST6/00400 (ZP) financed by the Polish National Science Center.

\appendix

\section{Moments of 3-tangle $\tau$} \label{sec:moments}

In the case of even moments of order $2 k$ of 3-tangle $\tau$
for a three qubit tensor it is easy to note, that it is
a linear combination of moments of rank $8^k$ for a random
normalized vector. The moments of normalized vectors
distributed uniformly can be calculated with a use of Beta integral.

Let $\vert \psi \rangle$ be a random unit vector of size $d$ distributed
uniformly on a complex sphere, then for a vector $p$ of non-negative integers
the following expectation value reads
\begin{equation}\label{eqn:beta-integral}
\langle  |\psi_1|^{2 p_1} |\psi_2|^{2 p_2} \dots |\psi_d|^{2 p_d}\rangle =
\Gamma(d) \frac{\Gamma(p_1+1) \Gamma(p_2+1)  \dots \Gamma(p_d+1) }{\Gamma(p_1+p_2+ \dots +p_d +d)}.
\end{equation}
One can also note, that all moments of random vector $\vert \psi \rangle$, which
are not in a form presented above, are equal to zero. This follows, for example,
form Collins-\'S{}niady formula~\cite{Col06} for integrals of
monomials over unitary matrices distributed with Haar measure.

Calculation of the second moment, $\langle \tau^2 \rangle = 8/55$,
is a simple task, but for higher moments the number of term grows so rapidly,
that we have used a package for symbolic computations \texttt{IntU},
which yield moments presented in Table~\ref{dtauta}. 
This package allows for exact calculation of polynomial integrals
over the unitary group with respect to the Haar measure~\cite{Puc11}.

\section{Bound for geometric measure of entanglement for tripartite states}

The law of Marchenko--Pastur describes asymptotic
behavior of singular values of non-hermitian, rectangular random matrices.
Let $X$ be a random matrix $X$ of size $N \times K$ with entries given by complex
random i.i.d. normal variables with zero mean and variance $1$.
We define $Y = X X^{\dagger}/ \mathrm{tr} X X^{\dagger}$ and 
for $c>1$ consider a random counting measure on a real line, which counts the number
of rescaled eigenvalues of $Y$ which belongs to a given set, i.e.
\begin{equation}
\mu_M(A) = \frac{1}{N} \# \{\lambda( c N Y) \in A\}.
\end{equation}
 
For a  measure defined above, if $N,K \to \infty$ with additional
assumption $K/N \to c$, there exist a limiting distribution $\mu_M \to \mu $ given by
\begin{equation}
d\mu(x) = \frac{1}{2 \pi} \frac{\sqrt{(a_{+} - x)(x - a_{+})}}{c x}  dx \text{ for } x\in [a_{-},a_{+}],
\end{equation}
with
\begin{equation}
a_{\pm} = ( 1 \pm \sqrt{c})^2.
\end{equation}
The above theorem gives us a behaviour of the largest eigenvalue of a matrix $Y$,
which is $\lambda_1(Y) \sim \frac{1}{c N}(1 + \sqrt{c})^2$.

In the case, when $N,K \to \infty$ but $K/N \to \infty$, the theorem does not give us
the limiting distribution, but form the theorem we will extract,
the rate of convergence of the largest eigenvalue.

Consider, the case when $K=N^2$, then $K/N = N$,
the direct usage of the Marchenko--Pastur law would give us a behaviour 
of the largest eigenvalue of $Y$, i.e. $\lambda_1(Y) \sim \frac{1}{N^2} (1 + \sqrt{N})^2 \sim \frac{1}{N}$ .

Now we will try to use the above asymptotics to bound the geometric measure of entanglement for a tripartite random states.

Consider a random tensor $\vert \psi \rangle \in \mathbb{C}^d \otimes
\mathbb{C}^d \otimes \mathbb{C}^d = \mathcal{H}_1 \otimes \mathcal{H}_2 \otimes \mathcal{H}_3$.
The geometric measure of entanglement is related to the overlap with the nearest product state, i.e.
\begin{equation}
\max_{\vert \phi  \rangle \in {\mathsf{sep}}} |\langle \psi \vert \phi \rangle |^2.
\end{equation}
In the above equation a set $\mathsf{sep}$ consist of vectors in a form 
$\phi_1 \otimes \phi_2 \otimes \phi_3$ for $\phi_i \in \mathcal{H}_i$.
For distinct $i,j,k \in \{1,2,3\}$, we denote $\mathsf{sep}_{i | j k}$ vectors of the form
$\phi_1 \otimes \phi_2$ for $\phi_1 \in \mathcal{H}_i$ and
$\phi_2 \in \mathcal{H}_j \otimes \mathcal{H}_k$.
We have, that  $\mathsf{sep} \subset \mathsf{sep}_{i | j k}$ which gives us
\begin{equation}
\max_{\vert \phi  \rangle \in {\mathsf{sep}}} |\langle \psi \vert \phi \rangle |^2
\leq
\max_{\vert \phi  \rangle \in {\mathsf{sep}_{i|j k}}} |\langle \psi \vert \phi \rangle|^2.
\end{equation}
The last maximum above is a square of the largest Schmidt coefficient for vector $\vert \psi \rangle$.

If we consider, the behaviour of a large random tripartite states the above inequality combined with
the relations obtained form the Marchenko--Pastur law we obtain 
\begin{equation}
\max_{\vert \phi  \rangle \in {\mathsf{sep}}} |\langle \psi \vert \phi \rangle |^2
\leq
\max_{\vert \phi  \rangle \in {\mathsf{sep}_{i|j k}}} |\langle \psi \vert \phi \rangle|^2 \sim \frac{1}{d}.
\end{equation}



%


%

\end{document}